\documentclass[a4paper,11pt]{article}

\usepackage{natbib}
\usepackage{eqnarray,amsmath}
\usepackage{amssymb}
\usepackage{graphicx}
\usepackage{tikz}

\usepackage{algorithm}
\usepackage{algorithmic}
\usepackage{subfigure}
\usepackage{bm}
\usepackage[margin=3cm]{geometry}
\usepackage{epstopdf}
\usepackage{authblk}
\usepackage{amsthm}

\theoremstyle{definition}

\newtheorem{remark}{Remark}

\title{\bf Improving the Accuracy of Marginal Approximations in Likelihood-Free Inference via Localisation}
\author[$\star$,$\dagger$]{Christopher Drovandi}
\author[$\ddagger$,$\mathsection$]{David J. Nott} 
\author[$\P$]{David T. Frazier} 

\affil[$\star$]{School of Mathematical Sciences, Queensland University of Technology, Australia} 
\affil[ ]{}
\affil[$\dagger$]{Centre for Data Science, Queensland University of Technology, Australia} 
\affil[ ]{}
\affil[$\ddagger$]{Department of Statistics and Data Science, National University of Singapore}
\affil[ ]{}
\affil[$\mathsection$]{Institute of Operations Research and Analytics, National University of Singapore}
\affil[ ]{}
\affil[$\P$]{Department of Econometrics and Business Statistics, Monash University, Australia}

\begin{document}

\newcommand{\vect}[1]{\boldsymbol{#1}}

\setlength{\parindent}{0pc}
\setlength{\parskip}{1ex}

\maketitle

\begin{abstract}
Likelihood-free methods are an essential tool for performing inference for implicit models which can be simulated from, but for which the corresponding likelihood is intractable.  However, common likelihood-free methods do not scale well to a large number of model parameters. A promising approach to high-dimensional likelihood-free inference involves estimating low-dimensional marginal posteriors by conditioning only on summary statistics believed to be informative for the low-dimensional component, and then combining the low-dimensional approximations in some way.  In this paper, we demonstrate that such low-dimensional approximations can be surprisingly poor in practice for seemingly intuitive
summary statistic choices.  We describe an
idealized low-dimensional summary statistic that is, in principle, suitable for marginal estimation.   
However, a direct approximation of the idealized choice
is difficult in practice.  We thus suggest an alternative approach to marginal estimation 
which is easier to implement and 
automate.  Given an initial choice of low-dimensional summary statistic that might only be informative about a marginal posterior location, the new method 
improves performance by first 
crudely localising the posterior approximation using all the summary statistics to ensure global identifiability, followed by a second step that hones 
in on an accurate low-dimensional approximation using the 
low-dimensional summary statistic.  
We show that the posterior this approach targets can be represented 
as a logarithmic pool of posterior distributions based on the 
low-dimensional and full summary statistics, respectively.  
The good performance of our method is illustrated in several examples.
\end{abstract}

\newpage

\section{Introduction}
\label{sec:intro}

Likelihood-free statistical methods such as approximate Bayesian computation (ABC, \citet{Sisson2018}) and Bayesian synthetic likelihood (BSL, \citet{Wood2010,Price2018}) are now commonly applied to conduct inference on the parameters of computationally expensive models for which simulation of synthetic data is easy, but likelihood computation is impractical.  Such approaches aim to find values of the model parameters for which simulated and observed data are close, typically on the basis of a set of summary statistics.

Standard likelihood-free techniques such as ABC and BSL are known to scale poorly to a large number of summary statistics \citep{Blum2009,frazier2020BSLasymp}. Although BSL methods scale more readily to higher-dimensional summaries \citep{priddle2019efficient}, its performance can still degrade as the dimension of the summaries increases. In spite of this, 
the use of a high-dimensional summary statistic might seem desirable for several reasons.  First, a high-dimensional summary statistic might be necessary to mitigate the inevitable information loss incurred by reducing the full dataset.  Second, when the model contains a large number of parameters, a large number of summary statistics are required by default: when the the dimension of the summary statistic is less than the parameter, it is not possible to point-identify all parameters.  
From a computational perspective, however, matching of simulated and observed summary statistics and posterior exploration become more difficult in higher dimensions. 

In the ABC framework, a promising approach for performing likelihood-free analyses for high-dimensional parameter problems is marginal
adjustment \citep{Nott2014} and the closely related copula ABC approach 
\citep{Li2017, chen+g19}.  
These approaches estimate low-dimensional (e.g.\ one or two dimensional) marginal posterior distributions, and then combine them in some way to obtain
an approximation to the joint posterior distribution.  The idea is that each parameter, or pair of parameters, is likely to be informed by only a small number of summary statistics.  Thus, in principle, it is possible to accurately estimate these low-dimensional posterior distributions, and avoid the curse of dimensionality associated with the use of high-dimensional summary statistics.  

The purpose of this paper is to demonstrate that the approximation of low-dimensional marginal posteriors based on intuitive low-dimensional summary statistics choices can be surprisingly poor in some cases, and less accurate than the corresponding approximation using all the summary statistics. 
Motivated by this observation, we describe a summary statistic choice for marginal posterior estimation that would, in principle, deliver precise marginal posterior inferences. Unfortunately, this statistic is not practically feasible to construct in the situations to which ABC is generally applied. Our more practical approach is based on crudely localising the posterior approximation using all the  summary statistics, and then subsequently conducting marginal inferences by matching low-dimensional summary statistics that are informative for different marginal parameters. We show that our two-stage approach can be thought of as sampling a logarithmic pool of posterior distributions based on low-dimensional and full summary statistics, respectively.

Although we focus on strategies for high-dimensional ABC based on combining low-dimensional marginal posterior estimates, there are various other techniques that have been employed to scale ABC methods to higher dimensions.  These include 
regression adjustment methods \citep{Beaumont2002,Blum2010} 
which can be used in conjunction with the methods we describe, and 
for which theoretical behaviour is explored in \cite{li+f18b}.  
Likelihood-free Gibbs or Metropolis-within-Gibbs approaches \citep{Kousathanas2016,clarte+rrs20,rodrigues+ns20} can break
down high-dimensional ABC problems into lower-dimensional ones, 
but these approaches can also increase the complexity of making good summary
statistic choices, similar to the case of marginal adjustment.   
Bayesian optimization approaches \citep{gutmann+c16} are particularly
useful for expensive simulators, and \cite{owen+psdkc20} consider 
a generalized Bayes method which is robust to model
misspecification and well-suited to high-dimensional problems. 
\cite{picchini+t22} consider guided proposals for sequential Monte Carlo
ABC schemes able to extend the applicability of these methods to
high-dimensional settings.  
A summary of earlier work on high-dimensional ABC methods is given by 
\cite{nott+ofs17}.  

Outside the ABC paradigm, there are many 
alternative approaches to likelihood-free
inference.  The focus of these works is not concerned with
high-dimensional problems explicitly, but they are better suited to this
than naive ABC approaches.  
Methods based on flexible conditional density estimation techniques applied
to estimation of a likelihood, likelihood ratios or the posterior density directly are popular;  
some of the available methods include random forests \citep{raynal+mprre18}, 
mixture or mixture of regression 
approaches \citep{bonassi+yw11,fan+ns13,papamakarios+m16}
neural density estimation techniques \citep{lueckmann+gbonm17,greenberg+nm19,papamakarios19,radev+mvak22} and 
density ratio estimation methods
using flexible classifiers \citep{cranmer+pl15,hermans+bl20,thomas+dckg21}.  Flexible estimation of likelihoods can be combined with flexible methods for
variational inference \citep{wiqvist+fp21,glockler+dm22}.   

The questions of summary statistic choice we address here are relevant
beyond the ABC setting, whenever summary statistics are used in 
likelihood-free inference and need
to be chosen with a focus on the approximation of low-dimensional
marginal posterior distributions.  Independently of the earlier ABC literature, 
\cite{benjamin+clw20}, \cite{jeffrey+w20} and \cite{miller+cflw21}
 also suggest that the estimation of low-dimensional 
 marginal posterior distributions or moments
can be easier than estimating the joint posterior distribution.  
ABC estimation of low-dimensional marginals using interpretable 
summary statistic choices may be particularly interesting 
when misspecification of the model is suspected.  
The idea of using insufficient summaries to discard information in misspecified
settings is recently discussed in \cite{Lewis2021}.
\cite{frazier+rr20} explore model misspecification in the ABC context and demonstrate that regression adjustment methods can
perform poorly compared to simple ABC methods.  Similar difficulties
are likely to occur with other regression or classification approaches, 
where a misspecified model
can result in observed summary statistics which are unusual for any
value of the model parameter, resulting in extrapolation 
beyond training data when computing approximations to the posterior
density.  In the case of neural likelihood-free inference 
methods, the effects of misspecification
have been explored recently by \cite{schmitt+bkr21}.
 
\section{Marginal Approximations in Likelihood-Free Inference} \label{sec:adj}

\subsection{Approximate Bayesian Computation}

We denote the parameter of the statistical model of interest as $\theta \in \Theta \subseteq \mathbb{R}^p$, where $p$ is the number of parameters.  The observed data is given by $y=(y_1,\dots,y_n) \in \mathcal{Y}^n$ where $n$ denotes the number of observations.  Ideally we wish to base our statistical inferences on the posterior density
\begin{align*}
\pi(\theta|y) \propto p(y|\theta)\pi(\theta),
\end{align*}  
where $p(y|\theta)$ is the likelihood function and $\pi(\theta)$ the prior density.  
However, for many complex models of interest, $p(y|\theta)$ may be too computationally expensive to permit the application of exact methods, up to Monte Carlo error, for approximating expectations with respect to the desired posterior.  In such situations, we can resort to likelihood-free methods, which replace likelihood evaluations with model simulations, to obtain an approximation to the posterior.  

Perhaps the most popular statistical likelihood-free method is ABC.  In ABC, we often firstly choose a summary statistic function, $S: \mathcal{Y}^n \rightarrow \mathcal{S}$ where $\mathcal{S} \subseteq \mathbb{R}^d$ and $d$ is the number of summary statistics, to map the data to a lower dimensional space.  We then compare observed data $y$ and simulated data $x \in \mathcal{Y}^n$ on the basis of the corresponding observed and simulated summary statistics. Overloading notation, we write the summary statistic function evaluated at a generic value for the data simply by $S$.  When it is necessary to distinguish between the summary statistic function evaluated at the observed data $y$ and simulated data $x$, we denote these by $S_y = S(y)$, and $S_x = S(x)$, respectively.  

 Write $\pi(\theta|S_y)\propto \pi(\theta)p(S_y|\theta)$ for the partial posterior distribution conditioned
 on the summary statistic $S_y$, where $p(S|\theta)$ is the summary statistic likelihood evaluated at $S\in\mathcal{S}$.  In ABC, the posterior density is approximated as follows:
\begin{align}
\pi(\theta|y) \approx \pi(\theta|S_y) \approx \pi_\epsilon(\theta|S_y) & \propto \pi(\theta) p_\epsilon(S_y|\theta), \label{eq:ABC_likelihood}
\end{align}
where $\pi_\epsilon(\theta|S_y)$ denotes the ABC posterior, and
\begin{align}
 p_\epsilon(S_y|\theta) = \int_{\mathcal{Y}^n}  p(S_x|\theta) K_\epsilon\{\rho(S_y,S_x)\} dS_x, \label{eq:likelihood_approx}
\end{align}
with $\rho: \mathcal{S} \times \mathcal{S} \rightarrow \mathbb{R}^+$ a distance function that compares observed and simulated data, and $K_\epsilon: \mathbb{R}^+ \rightarrow \mathbb{R}^+$ a kernel function that is designed to be relatively large when $\rho$ is small.  We will refer to $p_\epsilon(S_y|\theta)$ defined in \eqref{eq:likelihood_approx} as the ABC likelihood.  It is a kernel smoothed version of the true summary statistic likelihood $p(S_y|\theta)$.  Ultimately, for a given $S$ and $\rho$, the bandwidth parameter of the kernel $\epsilon$, also referred to as the ABC tolerance, defines when the observed and simulated data are considered close.  ABC can be more easily understood when using the indicator kernel function, $K_\epsilon\{\rho(S_y,S_x)\} = \mathbb{I}\{\rho(S_y,S_x) \le \epsilon\}$, which is equal to one when $\rho(S_y,S_x) \le \epsilon$ and 0 otherwise.  The integral in \eqref{eq:ABC_likelihood} for a given $\theta$ can be unbiasedly estimated by taking a single draw $x \sim p(\cdot|\theta)$ from the model and evaluating $K_\epsilon\{\rho(S_y,S_x)\}$.
 
Although there are many algorithms for sampling from the ABC posterior \citep{Sisson2018a}, in this paper we use the sequential Monte Carlo (SMC) ABC algorithm of \citet{Drovandi2011b}.  This method generates samples (often referred to as particles in the SMC context) from an adaptive sequence of ABC posteriors with decreasing ABC thresholds.  It does this by eliminating a proportion of particles with the highest discrepancy at each iteration.  The population of particles is rejuvenated via a resampling and move step, the latter achieved with an MCMC kernel to preserve the distribution of particles.  The number of MCMC iterations applied to each particle adapts with the overall MCMC acceptance rate.  In this paper we stop the SMC ABC algorithm when the acceptance rate in the MCMC step drops below $1\%$. 

\subsection{Marginal Approximations} \label{sec:marg_approx}

It is well known that simple ABC methods struggle to produce accurate approximations of the (partial) posterior as the dimension
of the summary statistic increases.  A Monte Carlo estimate
for the ABC likelihood \eqref{eq:likelihood_approx}  
results in a conditional kernel density estimator of the likelihood for the summaries, the statistical behavior of which is known to severely degrade as the dimension of the summary statistic increases, even
if the ABC tolerance is favourably chosen.  
Results in \cite{Blum2009}, \cite{Blum2010} and \cite{barber+vw15} make the effect
of the dimension on simple ABC algorithms more precise.  

Further theoretical
results for ABC methods are described in \cite{li+f18a} and \cite{frazier+mrr18}, and \cite{li+f18b} consider
regression adjusted ABC methods.  \cite{li+f18b,li+f18a} obtain interesting results
about both point estimation for the ABC posterior mean, and accurate 
uncertainty quantification of the ABC posterior, 
for rejection and importance sampling ABC
algorithms with and without regression adjustment for an appropriately chosen sample-size dependent tolerance.  With an adaptively chosen and informative proposal distribution, so long as the tolerance goes to zero fast enough, the authors demonstrate that ABC regression adjustment methods can control the Monte Carlo error of the posterior approximation and provide correct uncertainty quantification. However, it is not so easy to disentangle the impact of the summary statistic dimension in this theory,
since the construction of the required proposal becomes more
difficult as the parameter dimension increases, which is one situation 
when high-dimensional summary statistics are needed.  Furthermore, in the case of simple accept/reject ABC, results in \cite{frazier+mrr18} demonstrate that to control the Monte Carlo error  the tolerance used in ABC must be chosen as an increasing function of the summary statistic dimension. 
We note that, in the BSL framework, \cite{frazier2020BSLasymp} obtain similar results to
those of \cite{li+f18b} for regression adjusted ABC, showing that the two methods
behave similarly from a computational standpoint.  

An appealing approach for overcoming issues associated with likelihood-free methods in high dimensions is to approximate low-dimensional posterior distributions \citep{Nott2014,Li2017}.  Let $\theta_j$ be some component of $\theta$.  Note that $\theta_j$ may consist of more than one parameter, but for the purposes of this paper it suffices to treat $\theta_j$ as a scalar.  We note that $\theta_j$ may be informed by a relatively small number of summary statistics.  We denote the corresponding summary statistic function as $S_j: \mathcal{Y}^n \rightarrow \mathcal{S}_j$ where $\mathcal{S}_j \subseteq \mathcal{S}$, and the observed statistic as $S_{j,y} = S_j(y)$.  The motivation for this is that $\pi_{\epsilon_j}(\theta_j|S_{j,y})$ could be a much better approximation to $\pi(\theta_j|S_y)$ compared to $\pi_{\epsilon}(\theta_j|S_y)$ for two inter-related reasons: firstly, since $S_j$ is lower dimensional, it is easier to find matches between observed and simulated summary statistic vectors within a given tolerance region; secondly, since we are attempting to match summaries in a lower-dimensional space, we can more readily control the approximation error introduced through the use of a positive tolerance (i.e., it is generally the case that the tolerance value $\epsilon_j$ can be much smaller than $\epsilon$). For these reasons, there are meaningful benefits to employing low-dimensional summaries when targeting marginal inference in ABC.

In this paper we demonstrate that, even with an intuitively reasonable choice of $S_j$, sometimes the approximation $\pi_{\epsilon_j}(\theta_j|S_{j,y})$ can be less accurate than $\pi_{\epsilon}(\theta_j|S_y)$, and substantially so.  As an example of an intuitively reasonable choice, the semi-automatic ABC method of \cite{Fearnhead2012} produces summary statistics which are estimated posterior means for $\theta$, and $S_j$ could be obtained by extracting estimates for $\theta_j$.  That such an intuitive choice can perform badly is perhaps well-known, but it is worthwhile to demonstrate this in a simple normal example where using
sufficient statistics to produce posterior approximations may not produce 
accurate inferences for $\theta_j$.

\subsubsection{The Normal Case} \label{subsec:normal_example}
	Write $\mathcal{N}(\mu,\phi)$ for a normal distribution with mean $\mu$ and variance $\phi$, and write $\mathcal{N}(z;\mu,\phi)$ for the value of its density at $z$.  Let $y = (y_1,y_2,\ldots,y_n)$ be a random sample drawn from a $\mathcal{N}(\mu,\phi)$ distribution with unknown $\mu$ and $\phi$.  We assume that $\mu$ and $\phi$ are independent \emph{a priori} and allocate priors:  $\mu \sim \mathcal{N}(\mu_0,\phi_0)$ and $\phi \sim \mathcal{IG}(\alpha,\beta)$ where $\mathcal{IG}(\alpha,\beta)$ denotes the inverse-gamma distribution with shape $\alpha$ and scale $\beta$ and $\mu_0, \phi_0, \alpha, \beta$ are fixed.  A minimal and Bayes sufficient statistic for estimating $(\mu,\phi)$ jointly is $\bar{y} = \frac{1}{n} \sum_{i=1}^n y_i$ and $s^2 = \frac{1}{n}\sum_{i=1}^n(y_i - \bar{y})^2$, i.e.\ $\pi(\mu,\phi|y) = \pi(\mu,\phi|\bar{y},s^2)$.  Since $\bar{y}$ and $s^2$ are direct point estimates of $\mu$ and $\phi$, respectively, it might be tempting to consider approximating the marginals of these parameters conditioning only on their respective point estimate statistics.
	
\begin{figure}[!htp]
		\centering
		\subfigure[Marginal posterior densities for $\mu$]{\includegraphics[scale=0.5]{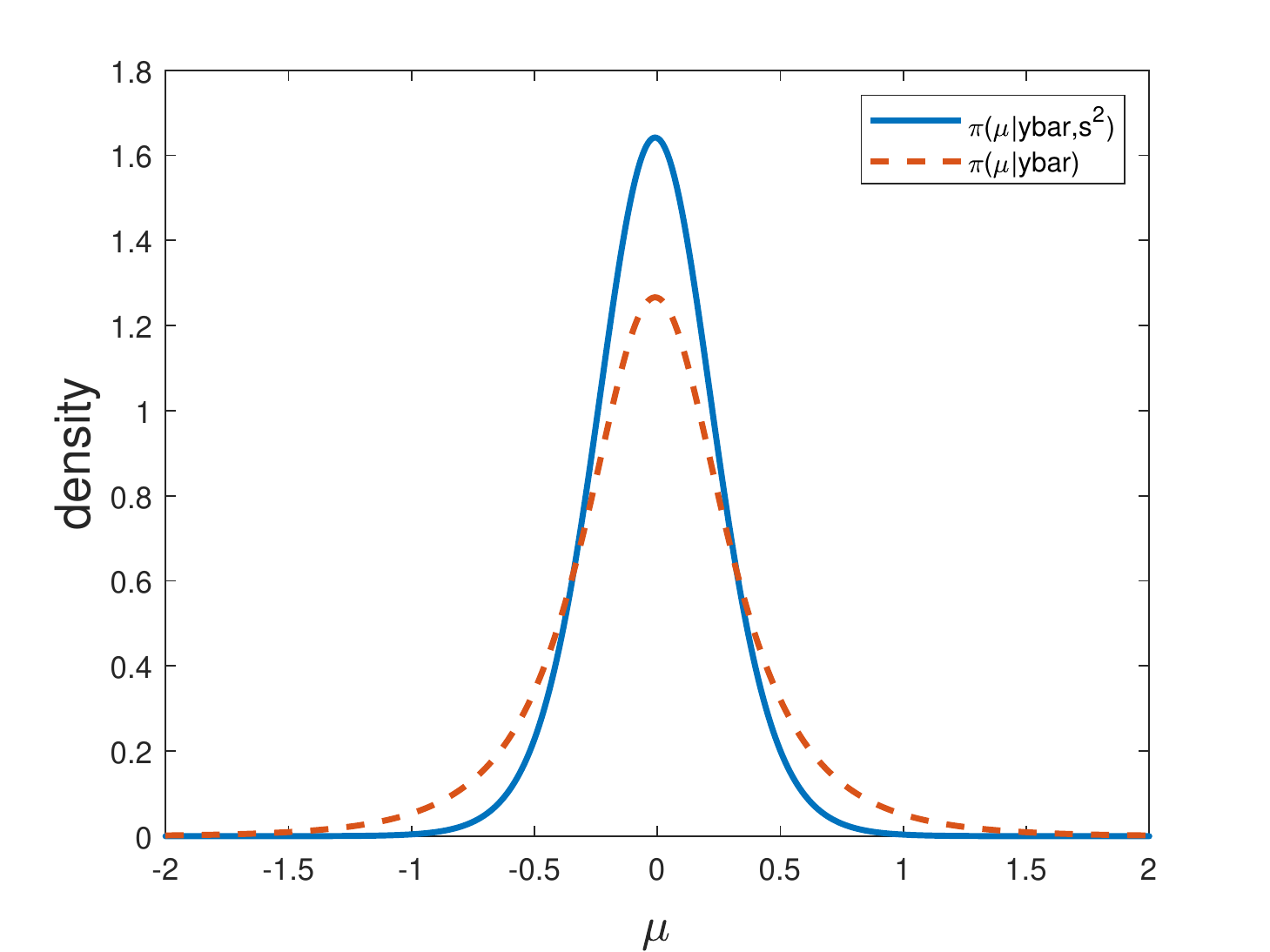}}
		\subfigure[Marginal posterior densities for $\phi$]{\includegraphics[scale=0.5]{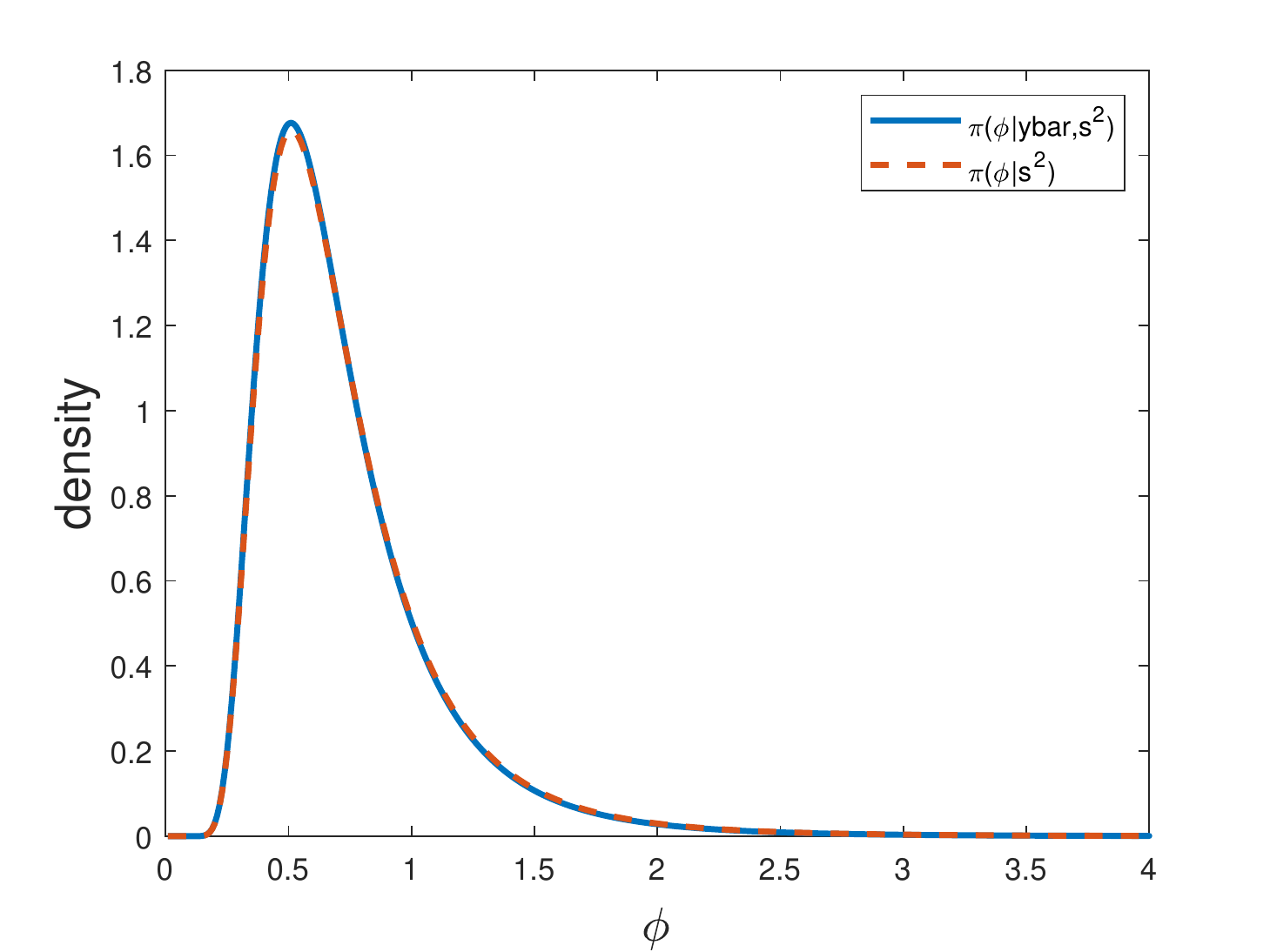}}
		\caption{Posteriors for toy normal example.}
		\label{fig:results_toy_normal}
	\end{figure}

	A certain realisation of $y$ with $n=10, \mu = 0, \phi = 1, \mu_0 = 0, \phi_0 = 1, \alpha = 1, \beta = 1$ produces the results shown in Figure \ref{fig:results_toy_normal}\footnote{Since most of the marginal posteriors do not have closed-form distribution, we normalise the densities using trapezoidal numerical integration for convenience, except for $\pi(\phi|s^2)$ which has an inverse-gamma distribution.}.  It is evident that $\pi(\mu|\bar{y})$ produces a diffuse approximation of the actual marginal posterior, whereas the approximation $\pi(\phi|s^2)$ is highly accurate for the actual $\phi$-marginal.

	To understand the findings in Figure \ref{fig:results_toy_normal}, it is helpful to first rewrite the normal likelihood as
	\begin{flalign*}
	\mathcal{N}(y;\mu,\phi)%
	&\propto \left\{\phi^{-(n-1)/2}\exp\{-s^2/2\phi\}\right\}\left\{\frac{1}{\sqrt{2\pi\phi}}\exp\{-n(\bar{y}-\mu)^2/2\phi\}\right\}\\&=p_1(s^2|\phi)p_2(\bar{y}|\mu,\phi).
	\end{flalign*}
	In the case of marginal inference on $\mu$, the above decomposition makes clear that marginal posterior inferences for $\mu$ will still depend on $s^2$. In particular, under the inverse-gamma prior for $\phi$ we know that 
	$$
	\pi(\mu|s^2,\bar{y})\propto \pi(\mu)\left[\beta+\frac{n(\mu-\bar{y})^{2}}{(n-1) s^{2}}\right]^{-\alpha-n / 2} .
	$$
	
	 If instead, we take as our ``likelihood'' for inference on $\mu$, $\exp\{-n(\bar{y}-\mu)^2/2\phi\}$,  then the marginal posterior for $\mu|\bar{y}$, under an inverse-gamma prior  for $\pi(\phi)$, is
	$$
	\pi(\mu|\bar{y})\propto \pi(\mu)\left[\beta+n(\mu-\bar{y})^2\right]^{-\alpha-1/2}.
	$$While the location of the two posteriors are similar, their scales are not, and the loss of precision that results from using the simpler likelihood increases posterior mass in the tails of the distribution, as evidenced in Figure \ref{fig:results_toy_normal}.
	
	Now we turn to the marginal approximation of $\phi$.  The term $p_2(\bar{y}|\mu,\phi)$ is non-constant for all $\phi$ except in the case where $\bar{y}=\mu$. Hence, we see that the \textit{likelihood contribution} of $p_2(\bar{y}|\mu,\phi)$ is non-constant in $\phi$, and hence \textit{is informative} for inference on $\phi$ \textit{based on this likelihood}; see, e.g., \cite{zhu1994information} for additional discussion on this example. 
	
	However, the marginal posterior for $\phi|s^2,\bar{y}$ and the marginal partial posterior for $\phi|s^2$ are given by
	\begin{flalign*}
	\pi(\phi|s^2,\bar{y})&\propto p_1(s^2|\phi)\pi(\phi)\frac{1}{\sqrt{2\pi\phi}}\int \exp\{-n(\mu-\bar{y})^2/2\phi\}\pi(\mu)d\mu,\\ \pi(\phi|s^2)&\propto p_1(s^2|\phi)\pi(\phi).
	\end{flalign*}
	The two posteriors appear different, however, in this particular case, the \textit{integrated term} $\pi(\phi|s^2,\bar{y})$ can be shown to be constant as a function of $\phi$ under a diffuse prior for $\mu$: if we have diffuse prior beliefs for $\mu$, considering the change of variables $z=\sqrt{n/2\phi}(\mu-\bar{y})$, we see that 
	\begin{flalign*}
	{(2\pi\phi)^{-1/2}}\int \exp\{-n(\mu-\bar{y})^2/2\phi\}\pi(\mu)d\mu&\propto {(2\pi\phi)^{-1/2}}\left(\frac{n}{2\phi}\right)^{-1/2}\int \exp(-z^2)dz\\&=\frac{1}{\sqrt{2\pi\phi}^{}}\left(\frac{\sqrt{2\pi\phi}}{n^{1/2}}\right)\\&=n^{-1/2},
	\end{flalign*}and thus we see that the integrated term in $\pi(\phi|s^2,\bar{y})$ is constant in $\phi$.

	Hence, even though the likelihood contribution containing $\bar{y}$, i.e., $p_2(\bar{y}|\mu,\phi)$, does carry meaningful information about $\phi$, the ``{likelihood}'' that results for integrating out $\mu$, under a diffuse prior, ensures that the integrated component $\int p_2(\bar{y}|\mu,\phi)\pi(\mu)d\mu$ has no information about $\phi$ that is not already contained in $p_1(s^2|\phi)$.  This explains the accurate results for $\phi$ in Figure \ref{fig:results_toy_normal}. In more substantive problems pertinent to likelihood-free inference, the integrated likelihood term that results from marginalization is unlikely to simplify in this manner, and we should not necessarily expect, \textit{a priori}, accurate marginal approximations, even for a seemingly obvious choice of low-dimensional summary statistic.
	
	On a more intuitive level, $s^2$ is obviously highly informative about $\phi$, and its distribution does not depend on $\mu$.  This might be the best situation that we can hope for when conditioning on a low-dimensional summary statistic consisting of point estimates of parameters in likelihood-free inference.  In contrast, even though $\bar{y}$ is informative about $\mu$, its distribution depends on $\phi$.
	
\subsubsection{A General Issue} \label{subsec:issue}
The normal example, while extremely simple, clearly illustrates the potential inaccuracy of using marginal approximations based on summary statistics which
are location estimates for the corresponding subset of parameters, and why the use of marginal approximations in ABC, based on such a choice of statistics, can produce inaccurate inferences
without careful summary statistic choice. Therefore, it is helpful to give a more
general formulation of the lessons of this example.

Consider that we wish to conduct inference on the unknown parameter $\theta=(\psi^\top,\lambda^\top)^\top$ conditional on a vector of observed summary statistics $S_{y}$, which we partition as $S_y=(S_{1,y}^\top,S_{2,y}^\top)^\top$, and which have dimension at least as large as $\theta$. Furthermore, consider the unrealistic scenario where the likelihood for the observed summary $S_y$ can be factorized as 
\begin{flalign}\label{eq:hopes}
p(S_y|\theta)\propto p_1(S_{1,y}|\psi)p_2(S_{2,y}|S_{1,y},\psi,\lambda).
\end{flalign}

The factorization in \eqref{eq:hopes} allows us to write the joint posterior as
$$
\pi(\theta|S_y)\propto \pi(\psi)\pi(\lambda)p_1(S_{1,y}|\psi)p_2(S_{2,y}|S_{1,y},\psi,\lambda).
$$Abusing notation, we can define an ``integrated likelihood'' for $\psi$ as 
$$p^\lambda_2(S_y|\psi):=\int p_2(S_{2,y}|S_{1,y},\psi,\lambda)\pi(\lambda)d\lambda,$$ 
and the marginal posterior $\pi(\psi|S_y)$ can then be written as
$$
\pi(\psi|S_y)=\frac{\pi(\psi)p_1(S_{1,y}|\psi)p^\lambda_2(S_{y}|\psi)}{\int p_1(S_{1,y}|\psi)p^\lambda_2(S_{y}|\psi)\pi(\psi)d\psi}.
$$

Critically, even in the unrealistic scenario where \eqref{eq:hopes} is valid, the marginal posterior $\pi(\psi|S_y)$ still depends on $S_{2,y}$ through the integrated likelihood term $p^\lambda_2(S_y|\psi)$. The only way that the posterior $\pi(\psi|S_y)$ does not depend on $S_{2,y}$ is when, for the observed value $S_y$, the term $p_2^\lambda(S_y|\psi)$ is a constant function for all values of $\psi$, i.e., when $p_2(S_{2,y}|S_{1,y},\psi,\lambda)$ does not depend on $\psi$. The latter requirement is essentially the requirement that the likelihood contribution $p_2(S_{2,y}|S_{1,y},\psi,\lambda)$ is $S$-nonformative (short for non-informative), a concept due to \cite{barndorff1976nonformation}, when conducting inference on $\psi$. 


Furthermore, the toy normal example makes clear that even in linear exponential families where the resulting decomposition in \eqref{eq:hopes} is satisfied, the likelihood $p_2(S_{2,y}|S_{1,y},\psi,\lambda)$ will in general depend on $\psi$. That is, while linear exponential families have minimally sufficient summaries, those same summaries are not generally $S$-nonformative when considering inference for a sub-vector of parameters, e.g., $\psi$. Given this finding, there is no reason to suspect that marginal approximations, by themselves, will yield accurate partial posterior inference for sub-vectors of parameters unless
the summary statistics are chosen in a very careful way.  The example in the next subsection demonstrates how such a careful summary selection can lead to accurate marginal approximations.  However, as we demonstrate in Section \ref{sec:examples}, finding summary statistics that satisfy the conditions outlined in this section is not an easy task. 

\subsubsection{Bivariate g-and-k example} \label{subsec:bivgandk}

The g-and-k distribution is a popular test example for ABC (e.g.\ \citet{drovandi2011likelihood} and \citet{Fearnhead2012}), and is defined by its quantile function
\begin{equation}
Q(z(p) | \theta) = a + b(1 + c \tanh[gz(p)/2]) (1+z(p)^2)^k z(p),  \label{eq:gandk}
\end{equation}
where $-\infty < a < \infty$, $b > 0$, $-\infty < g < \infty$, $k > -0.5$ and $z(p) = \Phi^{-1}(p)$ is the standard normal quantile function.  The parameters $\theta = (a, b, g, k)^\top$ control the location, scale, skewness and kurtosis respectively.  It is common practice to set $c = 0.8$.

\citet{drovandi2011likelihood} develop a multivariate (of dimension $J$) extension of the g-and-k distribution. A single realisation of dimension $M$, $x = (x^1,\ldots,x^M)$, from this distribution is generated by $(z(p)^1,z(p)^2,\ldots,z(p)^M) \sim \mathcal{N}_M(0,\Sigma)$ where $\Sigma$ is a correlation matrix and then by setting $x^m = Q^m(z(p)^m|\theta^m)$ for $m=1,\ldots,M$ where $Q^m$ is the quantile function for the $m$th marginal with corresponding g-and-k parameters denoted by $\theta^m = (a^m, b^m, g^m, k^m)^\top$. 

We consider here the bivariate g-and-k distribution with parameter   $\theta=({\theta^1}^\top,{\theta^2}^\top,r)^\top$ where $r$ is the single correlation parameter in $\Sigma$, which here is a $2 \times 2$ matrix.  For the observed data we generate $n=1000$ independent samples from the bivariate g-and-k model with parameter $\theta = (3, 1, 1, 0.5, 4, 0.5, 2, 0.5, 0.6)^\top$.  For the summary statistics we use robust measures of location, scale skewness and kurtosis (see Section \ref{subsec:gandk} for more details) for each marginal and the normal scores correlation coefficient \citep{Fisher1948}, which was also studied as the Gaussian rank correlation in \citet{Boudt2012}.  The latter summary statistic first computes the ranks for each of the data marginals, and then computes normal scores by feeding the scaled ranks into the standard normal quantile function.  Then the conventional sample correlation is computed between the scores.  Intuitively, the Gaussian rank correlation statistic is highly informative about $r$.  We consider running SMC ABC with 9 summary statistics, denoted $S$, and just the Gaussian rank correlation statistic, denoted $S_r$, to estimate the marginal posterior of $r$.

\begin{figure}[!htp]
	\centering
	\includegraphics[scale=0.6]{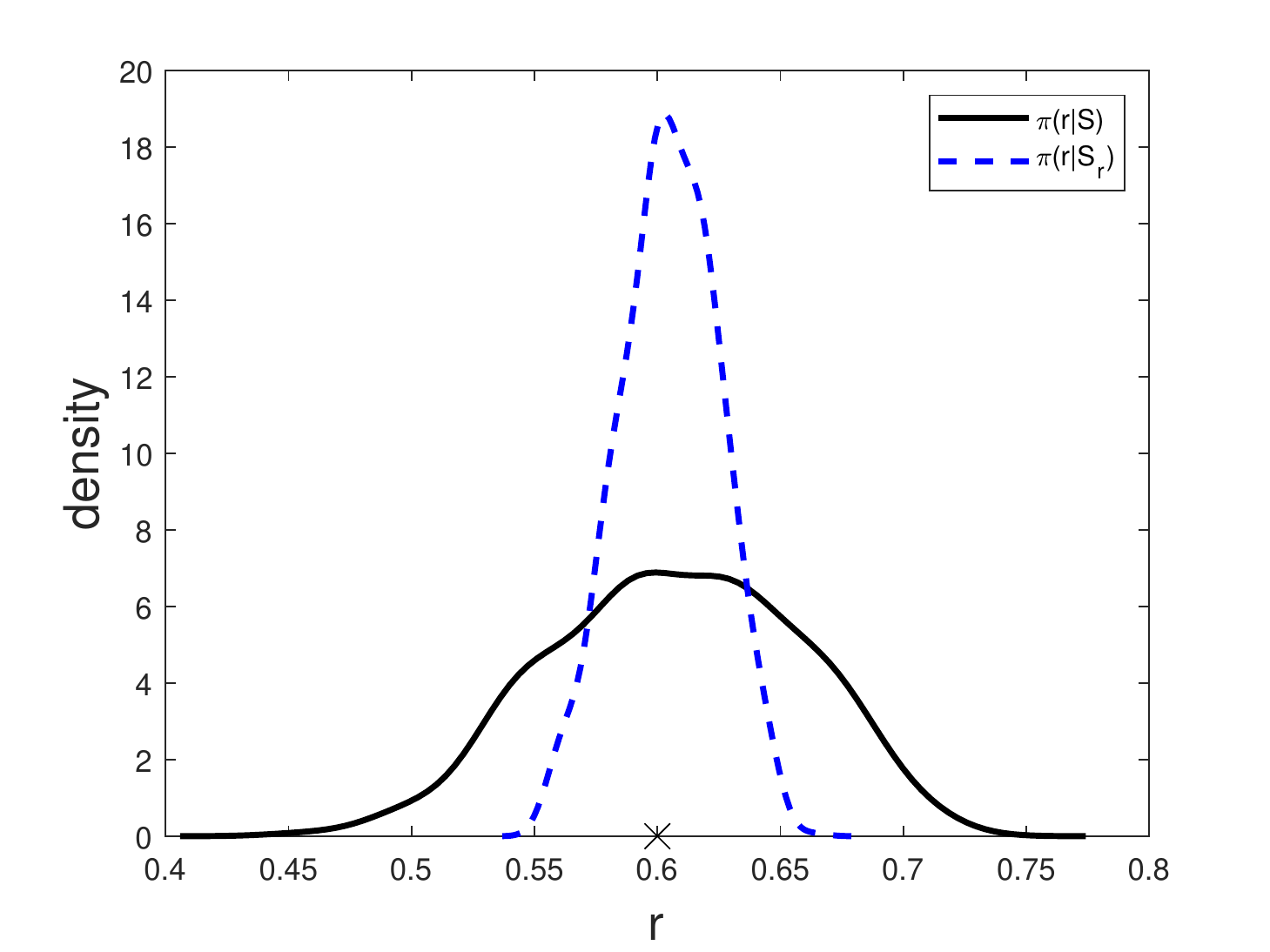}
	\caption{Approximate posteriors for $r$ for the bivariate g-and-k example.}
		\label{fig:results_bivgandk_rho}
\end{figure}

The results are shown in Figure \ref{fig:results_bivgandk_rho}. It is evident that a much more accurate marginal posterior approximation of $r$ can be achieved by conditioning on $S_r$ rather than the full set of summaries $S$.  We suggest the reason for this is that $S_r$ is non-parametric in the sense it uses the ranks of the marginal datasets rather than the raw values.  Since ranks are invariant to monotone transformations, $S_r$ is unaffected by the values of $\theta^1$ and $\theta^2$, that is the distribution of $S_r$ is independent of $\theta^1$ and $\theta^2$, i.e.\ $p(S_r|r, \theta^1, \theta^2) = p(S_r|r)$.  Combined with $S_r$ being highly informative about $r$ leads to an accurate approximation of the marginal posterior of $r$.

\section{Accurate marginal approximations via localisation}

In Appendix A we show that the idealized summary statistics for estimating the marginal partial posterior of $\theta_j$ are its corresponding marginal posterior moments.  For example, suppose we are interested in estimating the one-dimensional parameter $\theta_j$, and define an additional two-dimensional summary statistic vector $\widetilde{S}_j=(E(\theta_j|S),\text{Var}(\theta_j|S))^\top\in \widetilde{\mathcal{S}}_j$.  
Conforming to our previous notation, 
we write $\widetilde{S}_{x,j}$ and $\widetilde{S}_{y,j}$ for the value of $\widetilde{S}_j$ for simulated data $x$ and observed data $y$ respectively.   We show the posterior mean and variance of $\theta_j|\widetilde{S}_{y,j}$ is exactly the posterior mean and variance of $\theta_j|S_y$.  That is, we can match the mean and variance of the partial posterior with a much lower dimensional summary statistic, and this result extends to higher order moments.  However, we also explain in Appendix A why it is practically difficult to approximate these statistics in the likelihood-free setting, especially for higher-order moments.  Thus, as a more practical method, we propose the localisation approach described in the next section.

\subsection{Localisation approach}

We show how the issue describe in Section \ref{subsec:issue} can be mitigated by first localising the approximation using the full set of summary statistics $S$ and then focusing on matching the summary subset $S_j$.  More specifically, we consider the following approximate posterior for estimating the $\theta_j$-marginal
\begin{align}
\pi_{\epsilon_j}(\theta, S_x|S_y) & \propto \pi(\theta) p(S_x|\theta) \mathbb{I}\{\rho(S_y,S_x) \leq \epsilon_0\} \cdot \mathbb{I}\{\rho(S_{j,y},S_{j,x}) \leq \epsilon_j\}.  \label{joint-matching-main}
\end{align}
We abuse notation here and write $\rho(\cdot,\cdot)$ for the metric on both full and reduced summary statistic spaces. Here, the role of $\epsilon_0 > \epsilon_j$ is to provide an initial crude approximation of the posterior, so that the parameters are more tightly constrained compared to the prior.  Note that we do not use the data twice; in the second stage we simply impose a tighter constraint on $S_j$ that is designed to be informative about $\theta_j$.  Our approach is related to
other ABC methods using which use a pilot run to truncate the prior such as those of \cite{Fearnhead2012} and \cite{Blum2010}.  However, the key difference
in our approach is that different summary statistics are used at the pilot and 
analysis stage.

We now provide some intuition on why this idea is helpful for marginal approximations.  The joint selection condition is needed to ensure that values of $\theta$ are in the high density region for the approximation only if they produce reasonable global agreement for the entire vector of summaries; while the marginal selection condition for $S_j$ allows us to focus on specific regions of the marginal parameter space where $S_{j,y}$ is particularly informative about the the unknown $\theta_j$. However, in many cases the likelihood for the marginal summary statistic will be a complex function of all the parameters, and the joint selection step is critical as otherwise conducting inference using only the marginal selection condition could lead to diffuse posteriors, or, at worst, a complete lack of point identification. We later present examples of both types of behaviour, which emphasises the importance of including the joint selection condition when conducting marginal inferences. 

To implement this in practice we use a pilot run of SMC ABC with the full set of statistics $S$ for a relatively short time until the MCMC acceptance rate drops below some threshold much greater than $1\%$.   Then we perform a second SMC ABC step initialised at the pilot approximation and aim to produce closer matches based on $S_j$ (i.e.\ reduce $\epsilon_j$) until the MCMC acceptance rate drops below $1\%$.  Note that we also check in the second stage if a proposal satisfies the pilot constraint $\mathbb{I}\{\rho(S_y,S_x) \le \epsilon_0\}$.  If it does not, the proposal is rejected even if it matches the current constraint based on $S_j$.  It is important that the acceptance rate threshold for the pilot run is set relatively high.  Firstly, we do not want to use much computation time on the pilot run.  Secondly, in the continuation run, we want to reduce the tolerance associated with the marginal summary statistics greatly, and this will be difficult to do computationally if it is already difficult to match on the pilot tolerance.  Another possible stopping rule for the pilot run is when a certain number of model simulations have been exceeded, which may be a more explicit way to control the computational effort imposed in the pilot run.  The approach is summarised in Algorithm \ref{alg:method}.

\begin{algorithm}[tbh]
	\caption{{SMC ABC approach for accurate estimation of $\pi(\theta_j|S_y)$.} \label{alg:method} 
		\vspace{2mm}
		\newline
		{\em Inputs:} 
		Acceptance rate thresholds for the pilot, $p_{0,\mathrm{acc}}$, and continued, $p_{j,\mathrm{acc}}$, SMC ABC runs. Note that $p_{0,\mathrm{acc}} \gg p_{j,\mathrm{acc}}$.  Observed data $y$, prior distribution $\pi(\theta)$, discrepancy functions $\rho(S_y,S_x)$ and $\rho(S_{j,y},S_{j,x})$. 
		\vspace{2mm}
		\newline
		{\em Outputs:} Tolerances $\epsilon_0$ and $\epsilon_j$.  Samples $\{\theta^{(i)},\rho_j^{(i)},S_{j,x}^{(i)}\}_{i=1}^N$ from the approximate posterior $\propto \pi(\theta)p(S_x|\theta)\mathbb{I}\{\rho(S_y,S_x) \le \epsilon_0\} \cdot \mathbb{I}\{\rho(S_{j,y},S_{j,x}) \le \epsilon_j\}$.
	}
	\begin{algorithmic}[1]
		\STATE Run SMC ABC, reducing the tolerance with respect to the discrepancy function $\rho(S_y,S_x)$ until the acceptance rate falls below $p_{0,\mathrm{acc}}$.  This produces a sample $\{\theta^{(i)},\rho^{(i)},S_x^{(i)}\}_{i=1}^N$ from $\pi_{\epsilon_0}(\theta,S_x|S_y)$ where $\rho^{(i)} = \rho(S_y,S_x^{(i)})$ and $\epsilon_0 \geq \max \{\rho^{(i)}\}_{i=1}^N$.
		\STATE Compute $\rho_j^{(i)} = \rho(S_{j,y},S_{j,x}^{(i)})$ for $i = 1,\ldots,N$.  This produces the initial value of $\epsilon_j = \max \{\rho_j^{(i)}\}_{i=1}^N$.  We can now think of $\{\theta^{(i)},\rho^{(i)},S_x^{(i)}\}_{i=1}^N$ as an ABC sample from $\propto \pi(\theta)p(S_x|\theta)\mathbb{I}\{\rho(S_y,S_x) \le \epsilon_0\} \cdot \mathbb{I}\{\rho(S_{j,y},S_{j,x}) \le \epsilon_j\}$.
		\item Continue running SMC ABC, reducing the tolerance $\epsilon_j$ with respect to the discrepancy function $\rho(S_{j,y},S_{j,x})$, whilst still matching on $\mathbb{I}\{\rho(S_y,S_x)\le\epsilon_0\}$ for a fixed $\epsilon_0$, until the acceptance rate drops below $p_{j,\mathrm{acc}}$.  This produces samples $\{\theta^{(i)},\rho_j^{(i)},S_{j,x}^{(i)}\}_{i=1}^N$ based on the approximate posterior with kernel $\mathbb{I}\{\rho(S_y,S_x) \le \epsilon_0\} \cdot \mathbb{I}\{\rho(S_{j,y},S_{j,x}) \le \epsilon_j\}$.
	\end{algorithmic}
\end{algorithm}

\subsection{Interpretation in terms of logarithmic pooling}

The above approach has an interesting interpretation in terms of 
logarithmic pooling of posterior densities conditioned on summary statistic
vectors $S_y$ and $S_{j,y}$ respectively.  
For two densities $f_1(w)$ and $f_2(w)$ 
their log-pooled density with weight $\alpha$, $0\leq \alpha\leq 1$, is
$$f_p(w)=C(\alpha)f_1(w)^\alpha f_2(w)^{1-\alpha},$$
where $C(\alpha)$ is a normalizing constant, and we note that this definition can be extended
to accommodate any finite number of densities (\citealp{genest1986characterization}).  Logarithmic pools and other related pooling
methods are often used for constructing consensus priors when
prior distributions from multiple individuals are available, in the literature on optimal (Bayesian) decision making (see \citealp{genest1986combining} for a review) and in the literature on combinations of forecast distributions (see \citealp{clements2011combining} for a review).\footnote{In general, the weights for the logarithmic pool must be specified. This can be done in several ways, with the most common way being to obtain point estimates of the weights using data (\citealp{poole2000inference}). Alternatively, in certain settings, a prior distribution over the weights can be specified and a posterior distribution for the weights obtained via Bayes Theorem (see, e.g., \citealp{carvalho2022bayesian}). 
}

A key feature of logarithmic
pooling is that a high density value in the pooled density must necessarily have
a high density value in all the densities being pooled. We now argue that
that this feature can be used to obtain the joint matching condition in the SMC algorithm described above. Consider once again the 
ABC posterior density \eqref{eq:ABC_likelihood}.  Suppose we use the uniform
kernel, so that $K_\epsilon\{\rho(S_y,S_x)\} = \mathbb{I}\{\rho(S_y,S_x) \le \epsilon\}$.  Considering the summary statistic $S_y$, and denoting the tolerance
in our approximation by $\epsilon_0$, the density \eqref{density-joint}
is 
\begin{align}
  \pi_{\epsilon_0}(\theta,S_x|S_y) & \propto \pi(\theta) p(S_x|\theta) \mathbb{I}\{\rho(S_y,S_x)\le\epsilon_0\}. \label{density1}
\end{align}
Integrating out $S_x$ in \eqref{density1} gives the 
$\theta$-marginal density given by \eqref{eq:ABC_likelihood}.  
A similar ABC posterior density on $\theta$ and $S_x$ conditional on $S_{j,y}$, with tolerance denoted $\epsilon_j$ is
\begin{align}
  \pi_{\epsilon_j}(\theta,S_x|S_y) & \propto \pi(\theta) p(S_x|\theta) \mathbb{I}\{\rho(S_{j,y},S_{j,x})\le\epsilon_j\}. \label{density2}
\end{align}
 
It is immediate that for any $0<\alpha<1$, then for \eqref{density1} and \eqref{density2}, their log-pooled density is
\begin{align}
  \pi_{p,\epsilon}(\theta,S_x|S_y) \propto \pi(\theta)p(S_x|\theta)\mathbb{I}\{\rho(S_y,S_x) \le \epsilon_0\} \cdot \mathbb{I}\{\rho(S_{j,y},S_{j,x}) \le \epsilon_j\},  \label{logpool}
\end{align}
and the SMC-ABC algorithm described above targets sampling from
the density (\ref{logpool}), which involves the joint matching condition.  Note that
if we had integrated out $S_x$ first in \eqref{density1} and \eqref{density2} and
then pooled, this does not give the same result as pooling \eqref{density1} and \eqref{density2} and integrating out $S_x$ afterwards.  Based on the earlier interpretation of logarithmic pooling, a parameter value will only have high density under the pooled posterior if it has high density under the pilot ABC posterior and the ABC posterior where only $S_j$ is matched.  

The above makes it clear that, at least in the case of the uniform kernel, so long as $\alpha\in(0,1)$,  the weights in the logarithmic pool do not influence the resulting posterior.  The same result will not apply if one uses a bounded kernel function such as the triangular or Epanechnikov kernels. However, in the case of the commonly used Gaussian kernel, the effect of the pooling parameter $\alpha$ is just to change the effective kernel bandwidth parameters, since raising a Gaussian density to a power changes the scale, up to a normalizing factor.  Hence a joint matching condition for the Gaussian kernel also has an interpretation in terms
of logarithmic pooling.  

\section{Examples} \label{sec:examples}

The examples we consider only use a small number of summary statistics, so it is possible to use ABC to generate a gold standard approximation of the posterior distribution (and its corresponding marginals) conditional on these statistics.  This is intentional so that we can properly assess the performance of the various marginal approximations.  In the examples we also show the distribution of the simulated summary statistics for the accepted SMC-ABC posterior samples, so that we can compare how well the different approaches can match on individual summary statistics.  For simplicity we refer to such distributions as the posterior distributions of the summaries.

\subsection{MA(2) Example} \label{subsec:MA2}

Consider data $y = (y_1,y_2,\ldots,y_n)$ generated from the following moving average model of order 2 (MA(2)) model 
\begin{align}
y_{t}&=e_{t}+\theta _{1}e_{t-1}+\theta _{2}e_{t-2},  \label{MA2}
\end{align}
where, say, $e_{t}\sim \mathcal{N}(0,1)$ and the unknown parameters $\theta=(\theta_{1},\theta_{2})^\top$ are assumed to obey
\begin{align}
-2<\theta _{1}<2,\;\theta _{1}+\theta _{2}>-1,\theta _{1}-\theta _{2}<1.
\label{const1}
\end{align}
Our prior information on $\theta=(\theta_{1},\theta_{2})^\top$ is uniform over the invertibility region in \eqref{const1}. A useful choice of summary statistics for the MA(2) model are the
sample autocovariances $\eta _{j}(y)=\frac{1}{T}%
\sum_{t=1+j}^{T}y_{t}y_{t-j}$, for $j=0,1,2$.  Then we have $S_j = \eta_{j-1}$ for $j = 1,2,3$.   The \emph{binding functions}  corresponding to these summary functions (expected values of the summary statistics as a function of $\theta$) are given by
\begin{align*}
b_1(\theta) &= 1 + \theta_1^2 + \theta_2^2, \\
b_2(\theta) &= \theta_1 + \theta_1 \theta_2, \\
b_3(\theta) &= \theta_2.
\end{align*}
Since $\theta_1$ only appears in $b_1(\theta)$ and $b_2(\theta)$, it is tempting to approximate the marginal for $\theta_1$ matching only $S_1$ and $S_2$.  This turns out to be a poor choice generally as there are two solutions when we solve for $\theta_1$ and $\theta_2$ as a function of $b_1$ and $b_2$ in the binding function;  this suggests that the posterior conditional on $S_1$ and $S_2$ is likely to concentrate around two values in the parameter space.  Regarding estimation of the $\theta_2$-marginal, it may be sensible to match only on $S_3$.

We consider a dataset of size $n = 10000$ generated with $\theta_1 = 0.9$ and $\theta_2 = -0.05$.  When using only $S_1$ and $S_2$ as summary
statistics, there is a second value of $(\theta_1,\theta_2)$ matching
the value of $b_1$ and $b_2$ obtained for the true parameter value, $(\theta_1, \theta_2) \approx (0.4860, 0.7591)$, which is well separated from the true parameter.  We consider six ABC approximations: (1) matching all three summary statistics $S_1, S_2$ and $S_3$, (2) matching only on $S_1$ and $S_2$, (3) matching only on $S_3$, (4) a pilot ABC approximation matching $S_1, S_2$ and $S_3$, (5) continuing from the pilot approximation  matching only $S_1$ and $S_2$ and (6)  continuing from the pilot approximation  matching only $S_3$.   We stop the pilot run when the MCMC acceptance rate drops below $30\%$.  

\begin{figure}[!htp]
  	\centering
  	\includegraphics[scale=0.7]{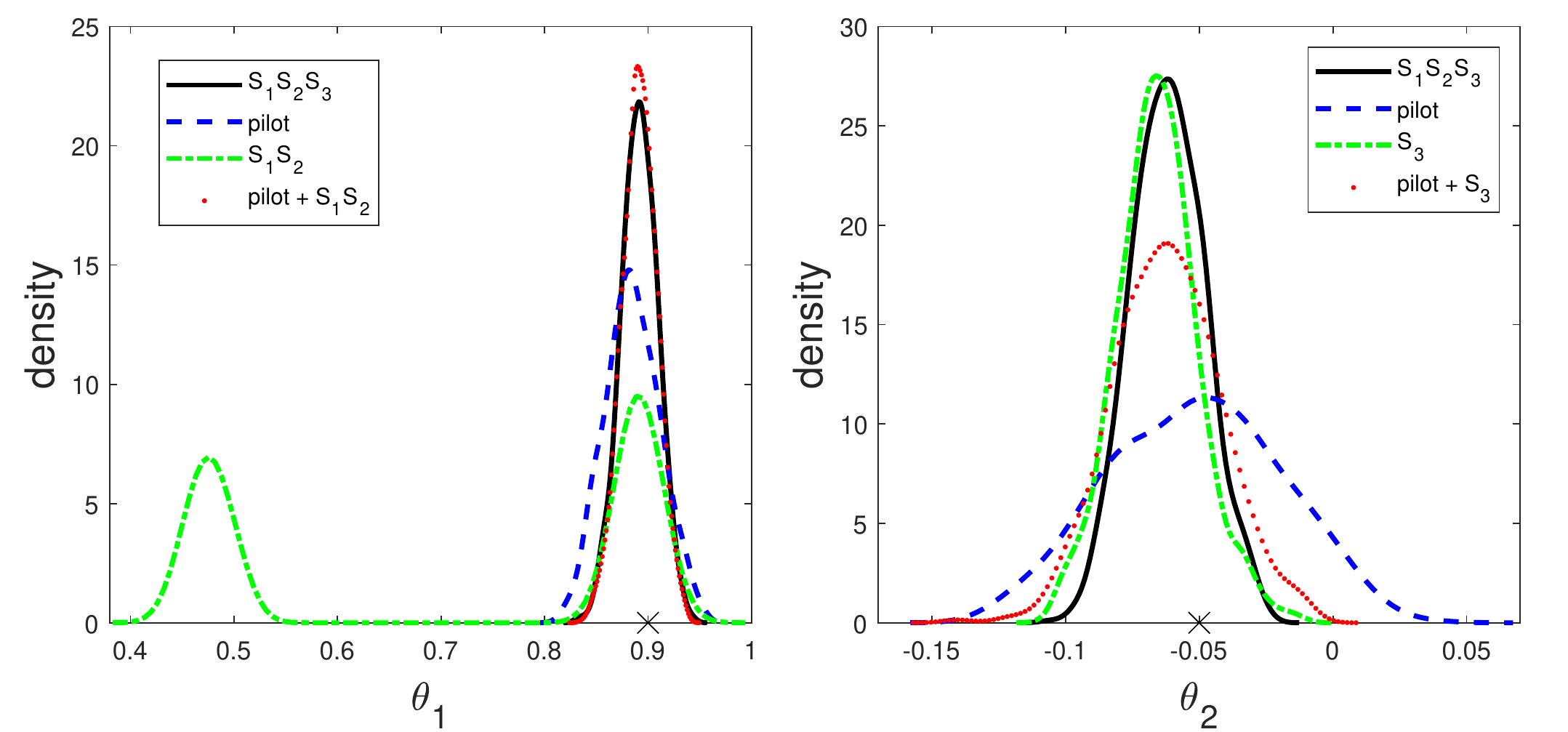}\label{figsub:results_parameters_ma2}
  	\caption{Univariate ABC posterior approximations for $\theta_1$ from various approaches for the MA(2) example.  Here $S_j$ indicates which statistic is used in the corresponding ABC approximation (the solid black lines are based on all three summary statistics),  ``pilot" indicates the pilot run with all statistics, and ``pilot + x" indicates the continuation of the pilot run with statistics x. The observed summary statistic is $S_y \approx (1.7999,    0.8363,   -0.0658)^\top$.  The $\times$ symbol on the x-axis represents the true value of the parameter.}
  	\label{fig:results_ma2}
  \end{figure}

  The results are shown in Figure \ref{fig:results_ma2}.  We first focus on estimating the $\theta_1$-marginal (left plot).  It can be seen that the marginal approximation of $\theta_1$ is poor when conditioning on only $S_1$ and $S_2$.  The posterior is multi-modal, as expected.  The pilot run (shown as blue dash) produces a diffuse approximation but effectively eliminates the second mode.  Then, continuing from the pilot run with $S_1$ and $S_2$ produces an accurate marginal approximation of $\theta_1$.

  Next we focus on the $\theta_2$-marginal (right plot).  This time, the standard marginal approximation based on only $S_3$ produces an accurate marginal approximation.  $S_3$ is a highly informative statistic for $\theta_2$, and the posterior for $\theta_1$ is close to the prior and shows little posterior dependence with $\theta_2$ (results not shown).  As we see with the $\theta_1$ marginal results and the examples below, obtaining an accurate marginal approximation without the pilot is an exception rather than the rule.  Here pilot $+ S_3$ gives a reasonable estimate of the $\theta_2$-marginal but it is not highly accurate. We suggest that this is a result on having to match on both the pilot and marginal tolerances, and so the pilot approximation may still have some influence.  We will see in the below examples that continuing from a pilot ABC run is crucial to obtaining accurate marginal approximations.

As alluded to above, the MA(2) example clearly demonstrates the identification issues that can arise when doing marginal adjustments in an ad-hoc fashion. Namely, since the distribution of the summaries $S_1,S_2$ results in a
multi-modal likelihood function as a consequence of the multiplicity of
roots discussed above, the information in these summaries is not enough by themselves to identify the true value of $\theta_1$. However, once we have adequately restricted the parameter space for $\theta_2$, which can be achieved through a pilot selection step based on all the summaries, the multiplicity of roots is alleviated, and the summaries $S_1,S_2$ are highly informative for the true value of the unknown parameter $\theta_1$. 

Since the parameter $\theta_2$ clearly influences the distribution of $S_1,S_2$, there is no hope of obtaining marginal sufficiency for $\theta_1$, when we base our inference for $\theta_1$ on the summaries $S_1,S_2$. Consequently, a joint selection step based on all the summaries will be required if we are to have any hope of identifying the unknown value of $\theta_1$. 

In contrast, the asymptotic distribution of the third summary statistic is (in the limit) independent of $\theta_1$, and hence inference for $\theta_2$ based solely on $S_3$ is likely to deliver reasonable results, as accords with the result in Figure \ref{fig:results_ma2}. In particular, we have that 
$$
S_{3,x}=n^{-1}\sum e_te_{t-2}=\theta_2n^{-1}\sum e_{t-2}^2+\theta_1n^{-1}\sum e_{t-1}e_{t-2}=\theta_2n^{-1}\sum e_{t-2}^2+o_p(1/\sqrt{n}),
$$	where the $o_p(1/\sqrt{n})$ comes about since $e_{t-1}$ is independent of $e_{t-2}$ for all $t$.\footnote{For $a_n\rightarrow\infty$ as $n\rightarrow\infty$, the notation $X_n=o_p(1/a_n)$ signifies that the random variable $a_nX_n$ converges to zero in probability.} Thus, for $n$ large $$S_{3,x}|\theta\sim\frac{\theta_2}{n}\chi^2(n),$$ which has mean $\theta_2$, variance $2\theta_2^2/n$, and does not depend on $\theta_1$.\footnote{From this result we also see that, for $n$ large,
$\sqrt{n}\{S_{3,x}-\theta_2\}|\theta\sim N(0,\theta^2_2)$ and the asymptotic distribution of the scaled and centred statistic only depends on $\theta_2$.} Hence, in large samples, 
\begin{flalign*}
p(S|\theta)&= p_1(S_{3}|\theta_1,\theta_2) p_2(S_1,S_2|\theta,S_3)\approx p_1(S_{3}|\theta_2) p_2(S_{1},S_{2}|\theta,S_3).
\end{flalign*}
Therefore, if we conduct inference on $\theta_2$ using just $S_3$ via the likelihood $p_1(S_3|\theta_2)$, the results are likely to be quite accurate, especially in the case where there is little information about $\theta_2$ in the conditional distribution $p_2(S_1,S_2|\theta,S_3)$. 

In general, the marginal posterior for $\theta_2|S$ in this example is given by 
$$
\pi(\theta_2|S)=\frac{\pi(\theta_2)p_1(S_3|\theta_2)p^{\theta_1}_2(S_1,S_2|\theta_2,S_3)}{\int\pi(\theta_2)p_1(S_3|\theta_2)p^{\theta_1}_2(S_1,S_2|\theta_2,S_3)d\theta_2}.
$$
In the extremal case where $p^{\theta_1}_2(S_1,S_2|\theta,S_3)$ is $S$-nonformative for $\theta_2$, the integrated likelihood is constant in $\theta_2$, and the posterior reduces to $\pi(\theta_2|S_3)\propto \pi(\theta_2)p_1(S_3|\theta_2)$. The latter posterior is plotted in Figure \ref{fig:results_ma2} as the green dotted curve, which we see is virtually identical to the marginal posterior $\pi(\theta_2|S)$ based on the full set of summaries. 

\subsection{Univariate g-and-k Example} \label{subsec:gandk}

Here we consider the univariate g-and-k distribution that is defined in Section \ref{subsec:bivgandk}.  As with previous studies, we consider analysing a simulated dataset of length $n = 10000$ with $a = 3$, $b = 1$, $g = 2$ and $k = 0.5$.  Writing $\mathcal{U}(a,b)$ for the uniform distribuiton on $[a,b]$, the prior on each component of $\theta$ is set as $\mathcal{U}(0,10)$ with independent components.

\citet{drovandi2011likelihood} consider using as summary statistics robust measures of location, scale, skewness and kurtosis, $S = (S_1,S_2,S_3,S_4)^\top$ with
\begin{equation*}
S_1 = L_{2}, \quad S_2 = L_{3}-L_{1}, \quad S_3 = \dfrac{L_{3}+L_{1}-2L_{2}}{S_2}, \quad S_4 = \dfrac{E_{7}-E_{5}+E_{3}-E_{1}}{S_2},
\end{equation*}
where $L_{i}$ denotes the $i$th quartile and $E_{i}$ denotes the $i$th octile.  Given the natural interpretation of the g-and-k parameters, it might be intuitively appealing for each component of $\theta$ to use the corresponding robust summary statistic.  That is, we use $S_1$, $S_2$, $S_3$ and $S_4$ for estimating the approximate posterior marginals of $a$, $b$, $g$ and $k$, respectively.  For the pilot run we use an MCMC acceptance rate threshold of $20\%$.

\begin{figure}[!htp]
	\centering
	\subfigure[Univariate posterior densities of parameters]{\includegraphics[scale=0.35]{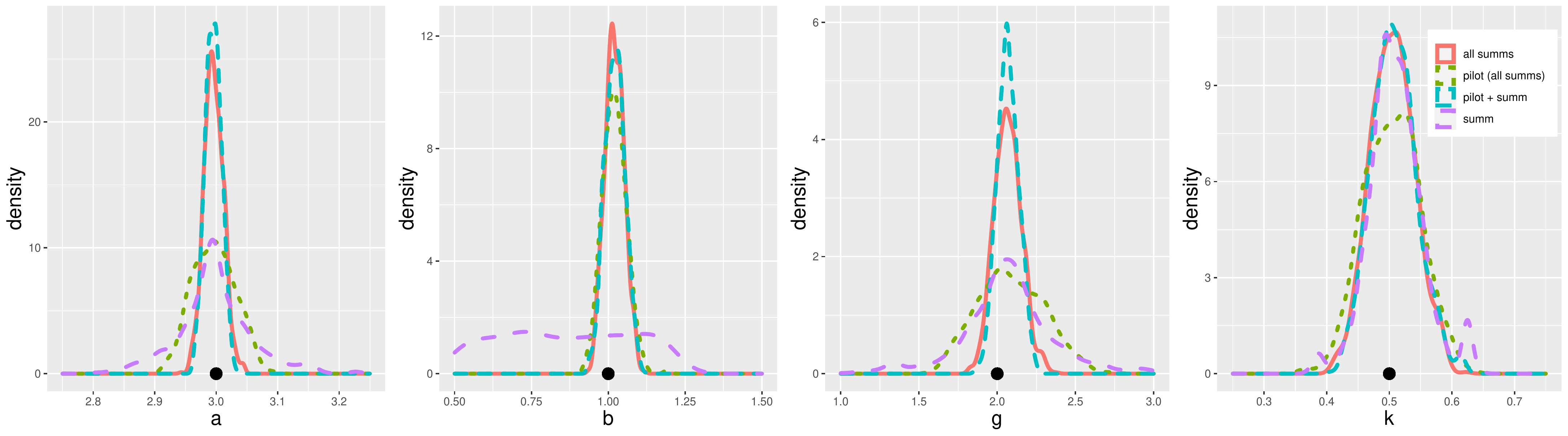}\label{figsub:results_gandk_parameters}}
	\subfigure[Univariate posterior densities of summaries]{\includegraphics[scale=0.35]{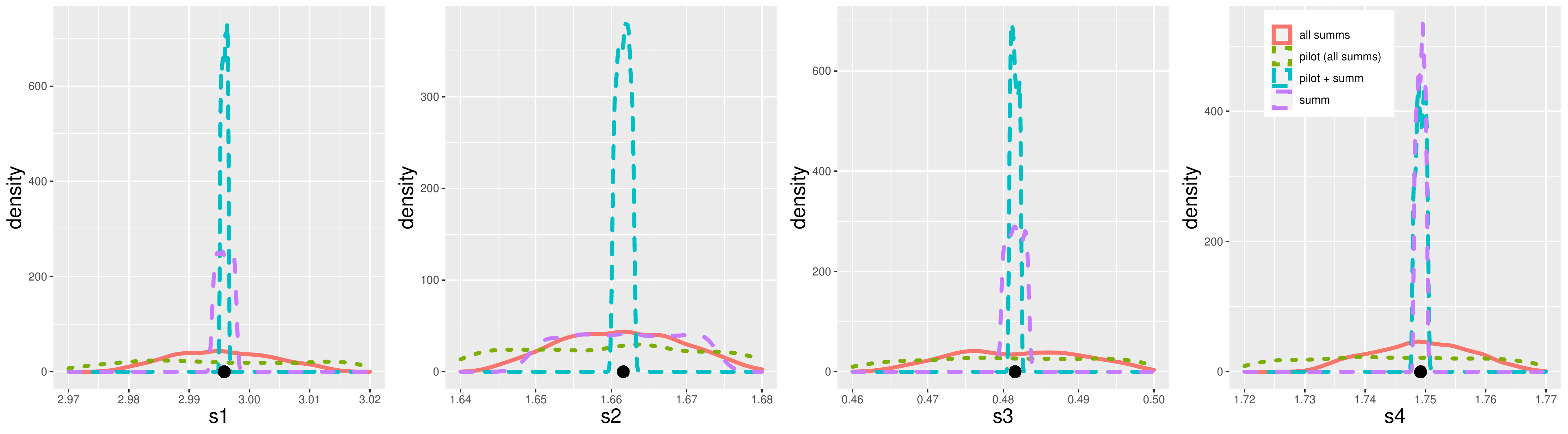}\label{figsub:results_gandk_summaries}}
	\caption{Univariate ABC posterior densities for the (a) parameters and (b) summary statistics obtained from various approaches for the g-and-k example.  Here ``all summs" means ABC with all four summary statistics, ``pilot (all summs)" means ABC with all four summary statistics but using a relatively large tolerance, ``summ" means ABC using the summary statistic relevant for its corresponding parameter and ``pilot + summ" means ABC using the summary statistic relevant for its corresponding parameter but also satisfying the tolerance from ``pilot (all summs)".  The dot on the x-axis shows the true parameter value in (a) and the observed summary statistic in (b).}
	\label{fig:results_gandk}
\end{figure}

The results are shown in Figure \ref{fig:results_gandk}.  Figure \ref{fig:results_gandk}a shows the univariate marginal parameter posterior approximations based on various approaches, whereas Figure \ref{fig:results_gandk}b shows the corresponding results for the four summary statistics.  The solid red densities in Figure \ref{fig:results_gandk}a show the typical ABC approximation based on all the statistics $S$.  Since there are only a small number of summary statistics, this approximation should be fairly close to $\pi(\theta|S_y)$.  The purple dash results show the marginal approximations when only using the $S_j$ statistic.  Apart from the parameter $k$, these marginal approximations are substantially worse than the marginal approximations based on $S$, despite producing closer matches to $S_j$ as demonstrated in Figure \ref{fig:results_gandk}b.  The pilot approximation results are shown in green dash.  Our approach using the pilot run as the initial approximation followed by close matching with $S_j$ produces the results shown as blue dash with univariate posterior approximations in agreement with the gold standard typical ABC approximation.  Figure  \ref{fig:results_gandk}b shows that this new approach results in very close matches with the $S_j$ statistics.

\begin{figure}[!htp]
	\centering
	\includegraphics[scale=0.5]{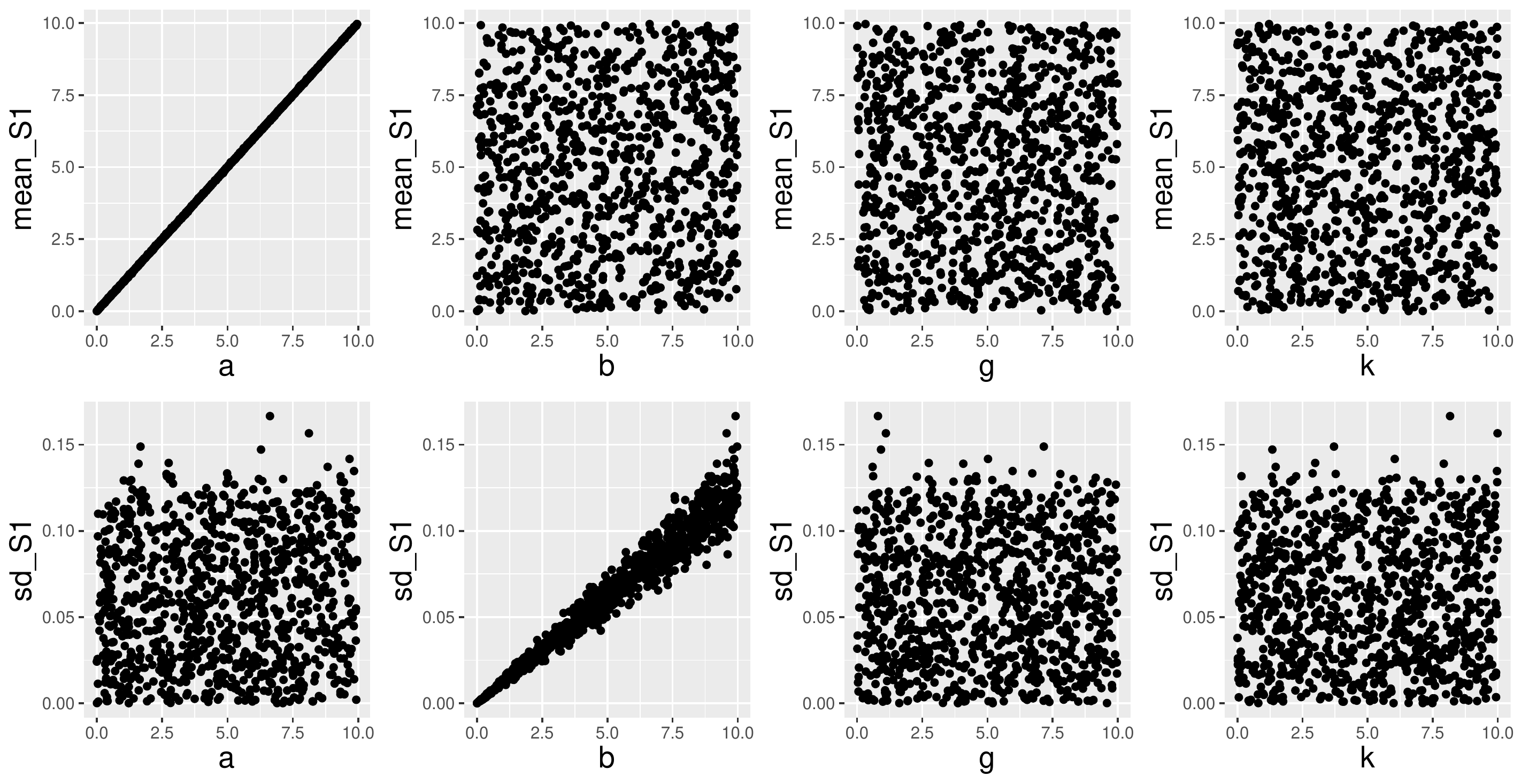}
	\caption{Scatterplots of the g-and-k parameters against the estimated mean (top row) and standard deviation (bottom row) of $S_1$.}
	\label{fig:examine_a_median_summary}
\end{figure}

The poor performance of the marginal approximation for $a$ based on only $S_1$ warrants further investigation, since the theoretical median of the g-and-k distribution is exactly $a$, and $S_1$ is the sample median.  One might anticipate the marginal approximation to work well in this case.   Here we examine the dependence of the distributional properties of $S_1$ on the parameters.  We do this by generating 50 independent replicates of datasets simulated based on 1000 draws from the prior predictive distribution.  For each individual dataset we compute $S_1$ the sample median.  For each of the 1000 prior predictive samples, we compute the sample mean and standard deviation over the 50 values of $S_1$.  Then we investigate how the mean and standard of $S_1$ is influenced by the g-and-k parameters.  Figure \ref{fig:examine_a_median_summary} shows scatterplots of the estimated mean and standard deviation of $S_1$ against the g-and-k parameters.  As expected, the parameter $a$ linearly influences the mean of $S_1$, and it appears that the other parameters do not affect the mean.  However, it is evident that the parameter $b$ strongly influences the standard deviation of $S_1$.  Therefore the distribution of $S_1$ depends on parameters other than $a$.

\begin{figure}[!htp]
	\centering
	\includegraphics[scale=0.4]{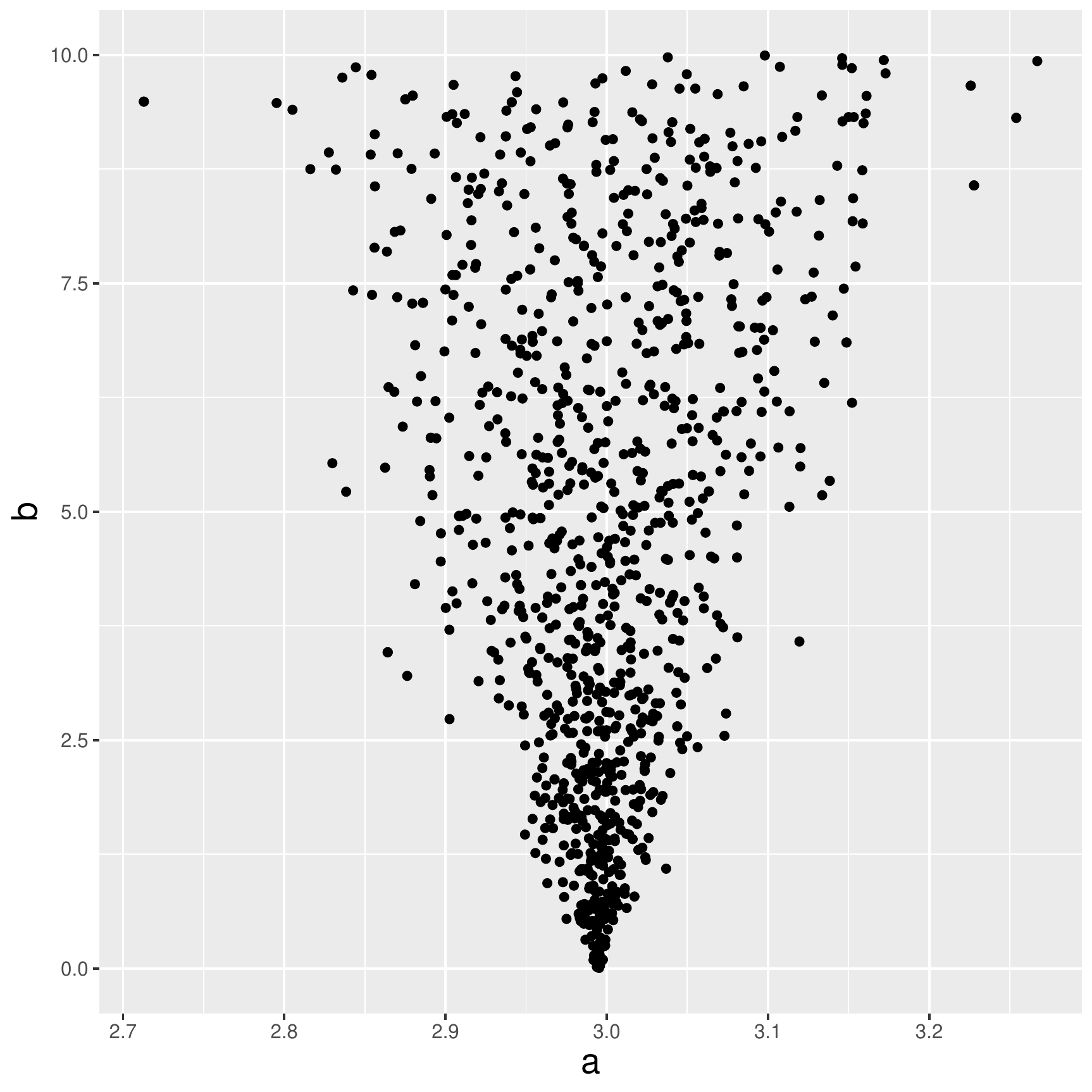}
	\caption{Scatterplot of the bivariate ABC posterior samples of $a$ and $b$ when conditioning only on $S_1$.}
	\label{fig:results_gandk_posterior_ab_S1}
\end{figure}

To further investigate the poor performance we show in Figure \ref{fig:results_gandk_posterior_ab_S1} a scatterplot of the bivariate ABC posterior samples of $a$ and $b$ when conditioning only on $S_1$.  It is evident that the posterior can support values of $a$ away from the true value of $3$ by making $b$ larger. Examining the form of the g-and-k quantile function in \eqref{eq:gandk}, by making $b$ larger we can make the simulated $S_1$ close to the observed $S_1$ just by chance even if the parameter $a$ is not very close to its true value.  The pilot run is able to discard the large values of $b$ by using information from all the statistics, allowing for an accurate marginal approximation of $a$ when continuing to match only on $S_1$.

We can again explain the link between $a$ and $b$ through the ideas discussed in Section \ref{sec:marg_approx}. Namely, in the case of the sample median statistic $S_1$, under the g-and-k distribution specification, the asymptotic distribution of $S_1$ is Gaussian with mean $a$ and variance $1/[n4\cdot f_{gk}(a|\theta)]$, where 	$f_{gk}(a|\theta)$ denotes the pdf of the g-and-k distribution, evaluated at the median $a$, and where the conditional notation $\theta$ clarifies that the pdf depends on the other parameters $b,g,k$. Consequently, even in large samples, it will not be the case that a decomposition like \eqref{eq:hopes} is satisfied for the summary $S_1$, 
and there is no hope that we can obtain a type of marginal sufficiency for $a$, even though $S_1$ is very informative about the mean behavior of $a$. 

While the pdf 	$f_{gk}(a|\theta)$  is in general intractable, setting $k=0$ and $g=0$, yields the normal distribution. In this case, the asymptotic distribution of $S_1|\theta$ is given by $$S_1|a,b,g=0,k=0\stackrel{a}{\sim}N[a,(\pi^2/2)(b^2/n)].$$Hence, even in the case of normal data, the distribution of $S_1$ depends on $b$, and this dependence can influence marginal inferences for the parameter $a$. 

The influence of $b$ on the marginal inferences of $a$ is precisely the culprit behind the marginal results for $a$ in Figure \ref{fig:examine_a_median_summary}, and the relationship between $(a,b)$ is precisely captured in Figure \ref{fig:results_gandk_posterior_ab_S1}. In particular, in large samples, the statistic $S_{1,x}$, in the case where $k=0=g$, can be written as 
$$
S_{1,x}=a+b\cdot\text{med}(z_1,\dots,z_n),\quad z_i\stackrel{iid}{\sim}N(0,1).
$$However, $\text{med}(z_1,\dots,z_n)\ne0$ for any finite $n$, and we can write $\gamma_n=\text{med}(z_1,\dots,z_n)$.

For marginal approximations based on $S_1$, we choose values of $a$ such that $|S_{1,x}-S_{1,y}|\le \epsilon$, for some tolerance $\epsilon$. Rewriting this condition as 
$$
-\epsilon \le a+b\gamma_n-S_{1,y}\le \epsilon
$$
clarifies that draws of $a$ are selected so long as  
$a+b\gamma_n-S_{1,y}\approx 0$. This can occur in at least two ways: one, when $b\gamma_n\approx0$, we select draws of $a$ such that $a\approx S_{1,y}$; two, when $b\gamma_n$ is large, we select draws of $a$ such that $a\approx b\gamma_n-S_{1,y}$. 
Since ABC is based on a joint simulation step for the parameters, values of $a,b$ that meet the second criterion above occur with non-zero probability, and are picked up by the algorithm simply because this combination of $(a,b)$ makes $|S_{1,x}-S_{1,y}|$ small. Given that  $\text{med}(z_1,\dots,z_n)$ is never zero in practice, a pure marginal selection step for $a$ produces posterior draws from both scenarios, with the latter scenario producing draws for $a$ that are far from its centre of posterior mass. However, as shown in Figure \ref{fig:examine_a_median_summary} the introduction of a pilot step mitigates this behavior by requiring reasonable values of $b$ in order for all the summaries to be matched.

\subsection{Robust Regression}

In this example we consider the setting of \citet{Lewis2021}, where interest is in performing Bayesian inference on a linear regression model of the form
\begin{align*}
y_i &= x_i^\top \beta + \epsilon_i, \mbox{ for } i = 1,\ldots,n,
\end{align*}
where $x_i$ are the covariates for the $i$th observation, $\beta$ are the regression coefficients (no intercept included) and $\epsilon_i$ are independent draws from a distribution with location 0 and scale $\sigma$.  Here the unknown parameter of interest is $\theta = (\beta, \sigma)^\top$.  

\citet{Lewis2021} are interested in generating robust Bayesian inferences for the linear regression model when the data contain outliers.  Instead of conditioning on the full dataset, \citet{Lewis2021} propose to condition on summary statistics that are robust to outliers, such as M-estimators.  \citet{Lewis2021} develop a method for exactly conditioning on such summary statistics, without having to resort to model simulation and likelihood-free inference.  However, \citet{drovandi+nf21} point out in their discussion of \citet{Lewis2021} that exact conditioning may be difficult to do for more complex regression models, and thus a likelihood-free approach may be appealing.

Here we use an insurance company dataset analysed in \citet{Lewis2021}.  The insurance company is interested in predicting the performance of insurance agencies, which is measured by the number of households the agency services (household count).  The data are grouped by states and we consider only state 27 here, which consists of data from 117 agencies.  The response variable is the square root of the household count for 2012 and the predictors are the square root of the household count from 2010 and two other covariates related to the size/experience measures of the number of employees associated with the agency (see \citet{Lewis2021} for more details).  The summary statistics are given by Huber's M-estimators of the regression coefficients and scale parameter, with the latter being log transformed here.  We label these statistics as $(S_1,S_2,S_3,S_4)$, each of which should be informative about $(\beta_1,\beta_2,\beta_3,\sigma)$, respectively. 

\begin{figure}[!htp]
	\centering
	\subfigure[univariate posterior densities of parameters]{\includegraphics[scale=0.4]{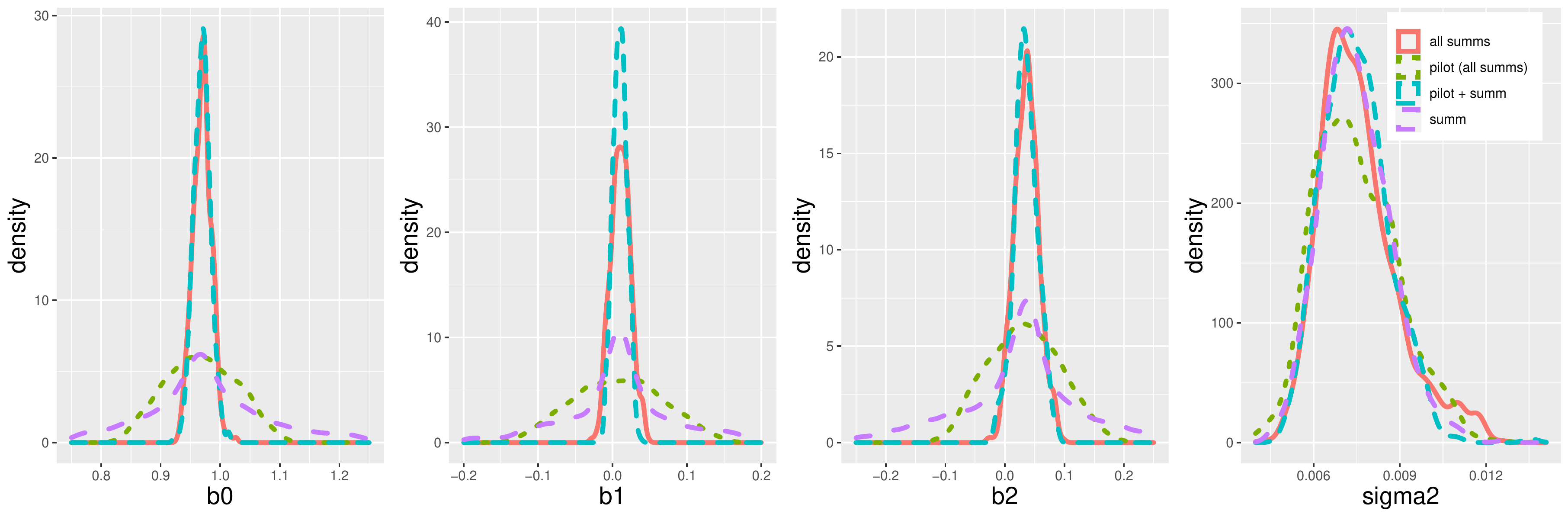}\label{figsub:results_regression_parameters}}
	\subfigure[univariate posterior densities of summaries]{\includegraphics[scale=0.4]{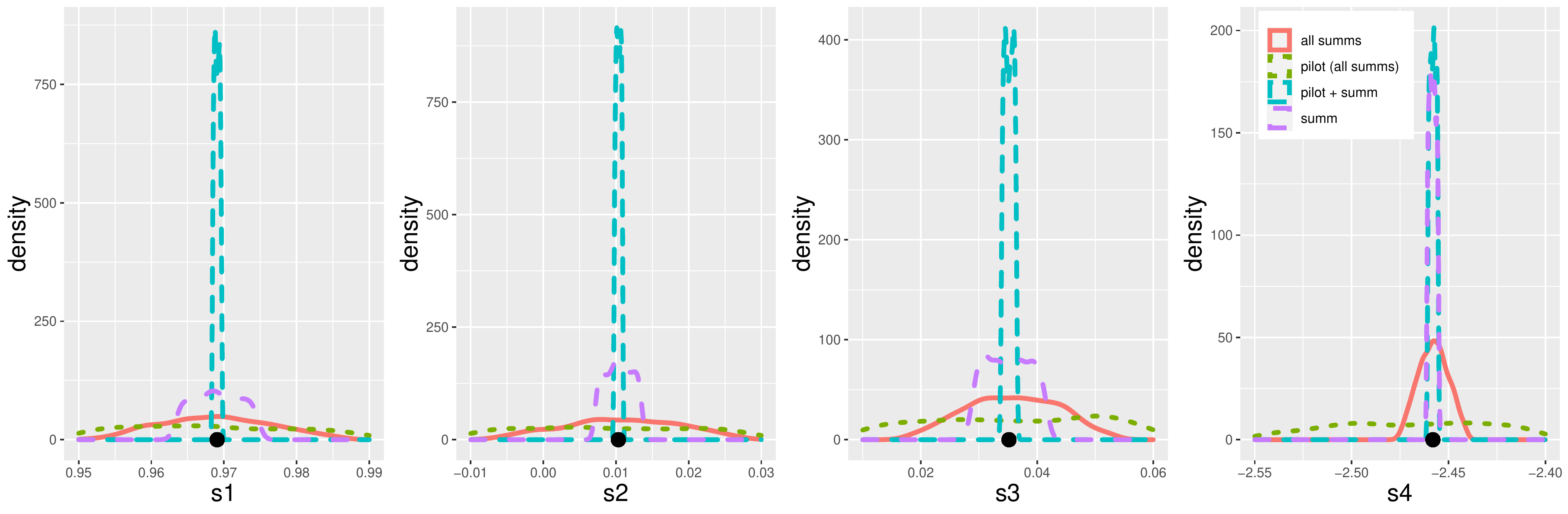}\label{figsub:results_regression_summaries}}
	\caption{Univariate ABC posterior densities for the (a) parameters and (b) summary statistics obtained from various approaches for the robust regression example.  Here ``all summs" means ABC with all four summary statistics, ``pilot (all summs)" means ABC with all four summary statistics but using a relatively large tolerance, ``summ" means ABC using the summary statistic relevant for its corresponding parameter and ``pilot + summ" means ABC using the summary statistic relevant for its corresponding parameter but also satisfying the tolerance from ``pilot (all summs)".  The dot on the x-axis in (b) is the observed value of the summary statistic.}
	\label{fig:results_regression}
\end{figure}

The results are shown in Figure \ref{fig:results_regression}.  As we can see, the results are qualitatively identical to the results of the g-and-k example.  The marginal approximations with each individual summary statistic are not accurate except for $\sigma^2$.  However, our localisation approach leads to accurate marginal posterior approximations.

\subsection{Lattice-Free Cell Model}

Collective cell spreading models are often used to gain insight into the biological mechanisms governing, for example, wound healing and skin cancer growth (e.g.\ \cite{VoDiameter2014,Vo2015}).   \citet{Browning2018} develop a simulation-based model where cells are able to move freely in continuous space.   Here we provide only brief details of the model and refer to \citet{Browning2018} for the full description.  Proliferation (cell birth) and motility (movement) for each cell evolves in continuous time according to a Poisson process.  The intrinsic rates are given by $p$ and $m$ for proliferation and motility events, respectively.  The rates of these processes are also neighbourhood-dependent, with rates decreasing as the amount of crowding around a cell increases.  The closeness of cells is governed by a Gaussian kernel that depends on a fixed cell diameter, $\sigma$.  When a cell proliferates, it places a new cell randomly in its neighbourhood according to an uncorrelated two dimensional Gaussian centered at the cell location with component variances of $\sigma^2$.  When a motility events occurs, the cell moves a distance of $\sigma$. The direction of the move depends on cell density, biased towards lower cell density.  A parameter used to help determine the move direction, $\gamma_b$, is part of a Gaussian kernel and measures the closeness of cells.  The parameter of interest is $\theta = (p,m,\gamma_b)^{\top}$.

In the experiments of \citet{Browning2018}, images of the cell population are taken every 12 hours starting at 0 hours with the final image taken at 36 hours.  \citet{Browning2018} use the number of cells and the pair correlation computed from each image as the summary statistics, resulting in a six dimensional summary statistic, $S$.  The pair correlation is the ratio of the number of pairs of agents separated by some pre-specified distance to an expected number of cells separated by the same distance if the cells were uniformly distributed in space.  The proliferation parameter $p$ is important as treatments would aim to reduce this parameter to slow tumour growth.  The number of cells at the end of the experiment should be highly informative about $p$.  Thus we consider marginal posterior approximations of $p$ by focussing on this statistic, which we label as $S_p$. 

The prior distribution is set as $p \sim \mathcal{U}(0,10)$, $m \sim \mathcal{U}(0,0.2)$ and $\gamma_b \sim \mathcal{U}(0,20)$ with no  dependence amongst parameters, as in \citet{Browning2018}.  Here we analyse a simulated dataset that is generated with true parameter value $\theta = (1,0.04,5)^\top$.

\begin{figure}[!htp]
	\centering
	\subfigure[univariate posteriors of parameters]{\includegraphics[scale=0.5]{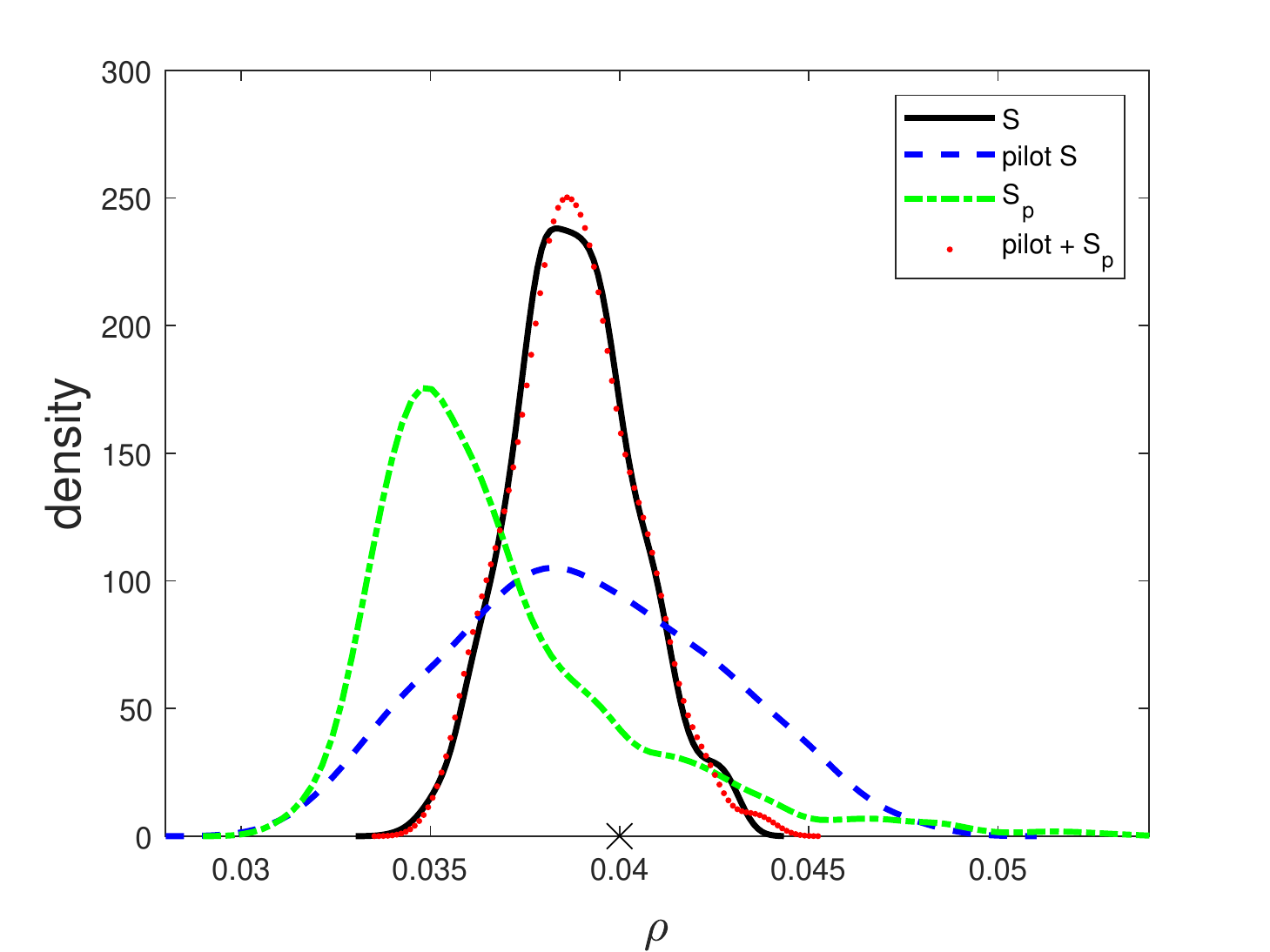}\label{figsub:results_latticefree_parameters}}
	\subfigure[univariate posteriors of summaries]{\includegraphics[scale=0.5]{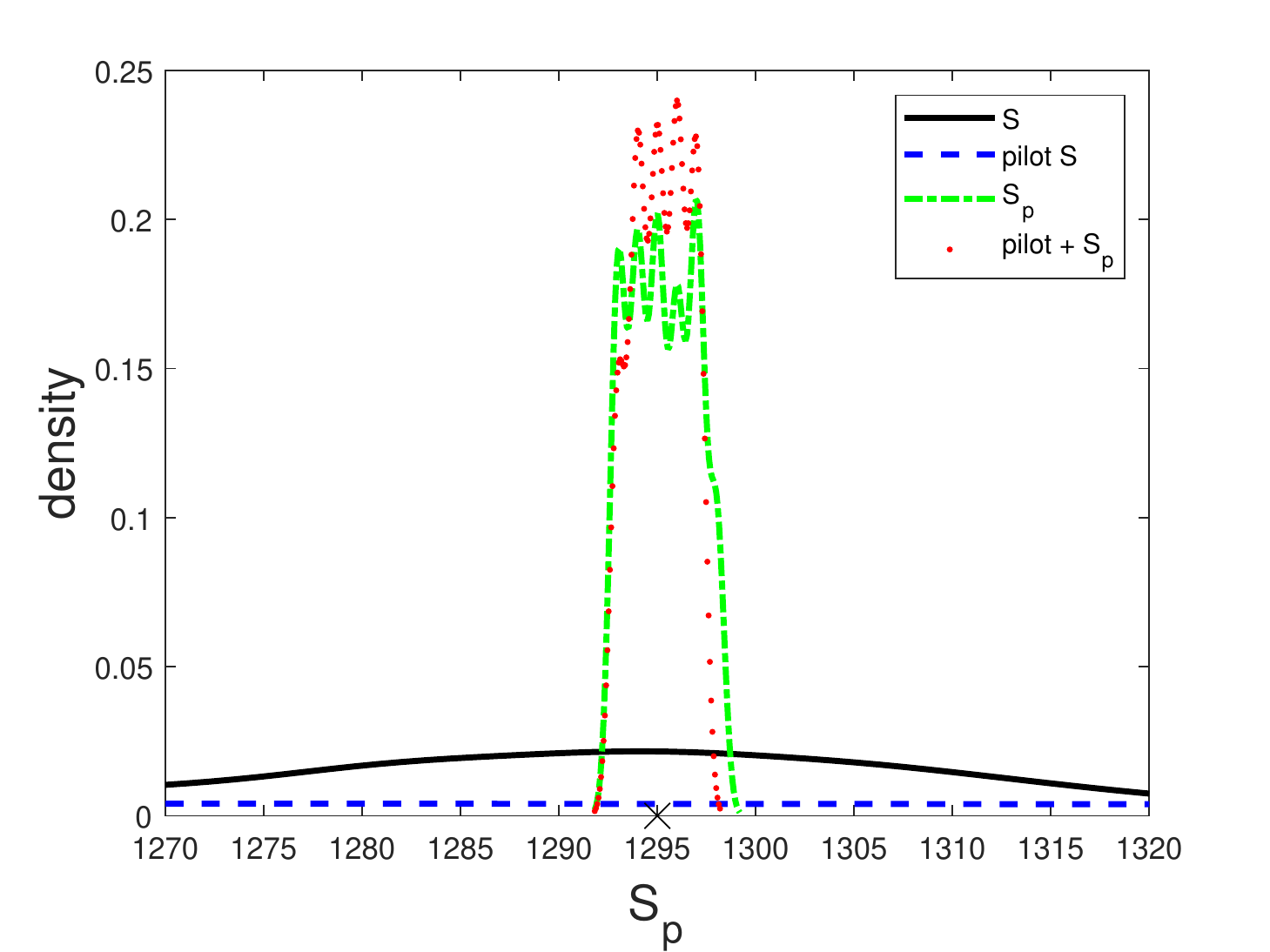}\label{figsub:results_latticefree_summaries}}
	\caption{Univariate ABC posterior distributions for the (a) $p$ and (b) $S_p$ from various approaches for the lattice-free cell example.  Here $S$ means ABC using $S$, ``pilot $S$" means ABC with $S$ but using a relatively large tolerance, $S_p$ means ABC using summary statistic $S_p$ only and ``pilot + $S_p$" means ABC using summary statistic $S_p$ but also satisfying the tolerance from ``pilot $S$".  The $\times$ symbol on the x-axis represents the true parameter value in (a) and the observed summary statistic in (b).}
	\label{fig:results_latticefree}
\end{figure}

The results are shown in Figure \ref{fig:results_latticefree}.  Since there are only 6 summary statistics, it is possible to get an accurate estimate of the partial posterior of $p$ by using standard SMC ABC matching on $S$, and we use that as the benchmark approximation (see the solid density in Figure \ref{figsub:results_latticefree_parameters}).   Then, we run SMC ABC matching only on $S_p$.  As evident from Figure \ref{figsub:results_latticefree_parameters}, this does not produce an accurate marginal approximation of $p$, despite generating closer matches to the observed $S_p$ compared to when running SMC ABC with $S$ (see Figure \ref{figsub:results_latticefree_summaries}).  Finally, we try our new approach.  Firstly, we perform a pilot run of SMC ABC using $S$ and stop the algorithm when the MCMC acceptance rate falls below 15\%.  Then, we continue from the pilot SMC ABC approximation by matching only on $S_p$.  As can be seen from  Figure \ref{figsub:results_latticefree_parameters}, this process produces an accurate approximation of the marginal posterior of $p$.

\section{Discussion} \label{sec:discussion}

In this paper we have presented a new approach for improving the accuracy of marginal approximations in likelihood-free inference when conditioning on a low-dimensional summary statistic.  We showed through several examples and sufficiency arguments that marginal approximations can be poor 
unless summary statistics are chosen very carefully, even if the summaries
are highly informative about the marginal posterior location. 
Our analysis demonstrates that, even if we have a vector of sufficient summaries, conducting ABC-based inference on a single (marginal) parameter using a well-intentioned subset of the summaries will not necessarily deliver inferences that agree with the partial posterior for the same (marginal) parameter based on the entire vector of summaries, even as $\epsilon\rightarrow0$. 

We have described some idealized summary statistics which are low-dimensional
and which can achieve accurate estimation of marginal posterior means and variances given the full set of summary statistics $S$ as the ABC tolerance
goes to zero. Given that such statistics may be difficult to approximate well, in practice, we find that joint matching
on the full summary statistics with a loose tolerance and  
low-dimensional summary statistics $S_j$ informative about
marginal posterior location with a much more stringent
tolerance results in greatly improved marginal
posterior estimation compared to matching using $S_j$ alone.  The joint matching
criterion can be motivated from the point of view of logarithmic pooling
of ABC approximations based on the full posterior and the summary statistics $S_j$.

Algorithmically, our approach used a crude pilot run of likelihood-free inference based on all the summary statistics in order to help identify the parameters, before honing in on the low-dimensional statistic intended to be informative for its corresponding low-dimensional parameter in a second stage. Our approach should combine well with the popular summary statistic selection approach of \citet{Fearnhead2012}, which forms summary statistics by estimating the posterior means via regressions.  \citet{Fearnhead2012} recommend performing a pilot run of ABC first in order to improve the fit of the regressions, which ultimately improves the derived summary statistics.  This pilot run can be exploited in our approach, which also requires a pilot run, and then the individual posterior mean estimates from the regression can be use as low-dimensional summaries for the corresponding low-dimensional parameters.

Our paper does not rigorously answer the question of how long the pilot ABC should be run.  In general, we suggest to run the pilot ABC until the acceptance rate starts to dramatically drop.  We want to perform the pilot run to eliminate very poor parts of the parameter space, but we do not want to run it so long that it becomes computationally difficult to satisfy the pilot discrepancy in the subsequent ABC run that focuses on matching on a low-dimensional summary statistic.   


In this article we used examples with a small number of summary statistics so that we could obtain a gold standard marginal approximation to compare our method against.  These examples were sufficient to illustrate the concepts and ideas of the paper.  However, we suggest that the principles of this paper can be extended to high-dimensional likelihood-free problems, and we leave this for future work.  
In truly high-dimensional problems, it may not suffice to crudely localise on
the full summary statistic $S$:  instead, and following the
discussion of Appendix A, methods will be needed to 
define subsets of $S$ which determine our idealized posterior mean
and variance statistics for each marginal for use in the pilot run.  
These subsets do not need to be very low-dimensional, but should not
be so high-dimensional that ABC methods are infeasible.  

\section*{Acknowledgements}

CD gratefully acknowledges support from the Australian Research Council Future Fellowship Award (FT210100260).  DTF gratefully acknowledges support by the Australian Research Council through grant DE200101070.

\bibliographystyle{apalike}
\bibliography{refs}

\newpage

\section*{Appendix A -- Idealized summary statistics for marginal estimation}

\subsection*{Definition of idealized summary statistics}

Can good summary statistics for marginal posterior estimation
be found?  We can answer this question affirmatively if the goal
is estimation of marginal posterior moments.  
For parameter $\theta_j$, we will define a certain
two-dimensional summary statistic vector, denoted by $\widetilde{S}_j$,
and then consider the marginal ABC posterior $\pi_{\epsilon_j}(\theta_j|\widetilde{S}_j)$.  We show that this marginal ABC posterior has
the same mean and variance as the exact partial posterior $\pi(\theta_j|S)$ 
as the tolerance $\epsilon_j\rightarrow 0$.  The 
idealized summary statistics we describe are not easily computable;  practical issues are discussed later.  Our suggestion extends 
the posterior mean summary statistics
$E(\theta|S)$ of \cite{Fearnhead2012}, which are optimal for ABC point estimation
in a certain sense, and which are also not directly computable without further approximation.   

To motivate the proposed summary statistics, we first note that the ABC posterior (\ref{eq:ABC_likelihood}) is the $\theta$-marginal density of a certain joint density on $\theta$ and a replicate summary statistic value $S_x$:
\begin{align}
\pi_{\epsilon}(\theta,S_x|S_y) & = \frac{\pi(\theta) p(S_x|\theta) \mathbb{I}\{\rho(S_y,S_x) \le \epsilon\}}{\int\int\pi(\theta) p(S_x|\theta) \mathbb{I}\{\rho(S_y,S_x)\le \epsilon\}d\theta dS_x}, \label{density-joint}
\end{align}where integrating out $S_x$ in \eqref{density-joint} gives the 
$\theta$-marginal density in \eqref{eq:ABC_likelihood}.   Write ${\cal B}_\epsilon$ for the set $\{S_x:\rho(S_x,S_y)\le\epsilon\}$, and for $S_x\in\mathcal{B}_\epsilon$, we can write 
$$p_x(S_x|\mathcal{B}_\epsilon)=\frac{\int\pi(\theta)p(S_x|\theta)d\theta\mathbb{I}(\mathcal{B}_\epsilon)}{\int\int\pi(\theta)p(S_x|\theta)d\theta\mathbb{I}(\mathcal{B}_\epsilon)d S_x}$$ for the ``restricted marginal density'' of $S_x$ in \eqref{density-joint}, which, by construction, is restricted to have support $\mathcal{B}_\epsilon$. 
From $p_x(S_x|\mathcal{B}_\epsilon)$, we see that the posterior in \eqref{density-joint} can be rewritten as 
\begin{align}
\pi_\epsilon(\theta,S_x|S_y) = \pi(\theta|S_x)p_x(S_x|\mathcal{B}_\epsilon) \label{joint-rep}
\end{align}Note that, when the summary $S$ is such that $S=S_x$, and if $S_x\in\mathcal{B}_\epsilon$, then the conditional density $\pi(\theta|S_x)$ in \eqref{joint-rep} is the exact partial posterior density conditional on $S$, irrespective of the value of $\epsilon$.

Now, suppose we are interested in estimating the one-dimensional parameter $\theta_j$, and define an additional two-dimensional summary statistic vector $\widetilde{S}_j=(E(\theta_j|S),\text{Var}(\theta_j|S))^\top\in \widetilde{\mathcal{S}}_j$.  
Extending our previous notation, 
we write $\widetilde{S}_{x,j}$ and $\widetilde{S}_{y,j}$ for the value of $\widetilde{S}_j$ for simulated data $x$ and observed data $y$ respectively.  
Consider the ABC posterior 
\eqref{density-joint} for $(\theta,S_x)$, but where the selection step is carried out using only $\widetilde{S}_j=\widetilde{S}_{y,j}$ rather than $S=S_y$.  
Using the representation \eqref{joint-rep}, and for $\widetilde{\mathcal{B}}_\epsilon:=\{S_x:\rho(\widetilde{S}_{j,x},\widetilde{S}_{j,y})\le\epsilon\}$, we then have
\begin{align}
\pi_\epsilon(\theta,S_x|\widetilde{S}_{y,j}) & = \pi(\theta|S_x)p_x(S_x|\widetilde{\cal B}_\epsilon). \label{abc-posterior-subset}
\end{align}
We assume that $\rho(\cdot,\cdot)$ satisfies the condition that for some 
constant $C>0$
$$C\|x-y\|\le \rho(x,y),$$
for all $x,y\in \widetilde{\mathcal{S}}_j$, where $\|\cdot\|$ is the Euclidean norm, 
a condition that holds for most commonly used ABC distance measures.  
This implies that if 
$S_x\in \widetilde{\mathcal{B}_\epsilon}$, then 
$$C \|\widetilde{S}_{x,j}-\widetilde{S}_{y,j}\|\le\epsilon,$$ 
The term on the left is greater than both $C|E(\theta_j|S_x)-E(\theta_j|S_y)|$ and 
$C|\text{Var}(\theta_j|S_x)-\text{Var}(\theta_j|S_y)|$.  Hence 
if $S_x\in \widetilde{\mathcal{B}}_\epsilon$ then
\begin{align}
& |E(\theta_j|S_x)-E(\theta_j|S_y)|\le C^{-1}\epsilon\;\;\;\text{ and }\;\;\;|\text{Var}(\theta_j|S_x)-\text{Var}(\theta_j|S_y)|\le C^{-1}\epsilon.  \label{bounds}
\end{align}

Writing $E_\epsilon(\cdot)$ for an expectation 
with respect to $\pi_\epsilon(\theta,S_x|\widetilde{S}_{y,j})$, we see that
\begin{align*}
E_\epsilon(\theta_j) & = E_\epsilon(E(\theta_j|S_x))  = \int E(\theta_j|S_x) p_x(S_x|\widetilde{\mathcal{B}}_\epsilon)\,dS_x
\end{align*}
which, by (\ref{bounds}) and dominated convergence, implies \
\begin{align}
\lim_{\epsilon\rightarrow 0}E_\epsilon(\theta_j) & = E(\theta_j|S_y).  \label{moment1}
\end{align} 

Writing $\text{Var}_\epsilon(\cdot)$ for a variance computed for the
joint density $\pi_\epsilon(\theta,S_x|\widetilde{S}_{y,j})$, and using
the law of total variance, 
\begin{align}
\text{Var}_\epsilon(\theta_j) & = E_\epsilon(\text{Var}(\theta_j|S_x))+
\text{Var}_\epsilon(E(\theta_j|S_x)). \label{law-of-tv}
\end{align}

Using \eqref{bounds} and dominated convergence, 
the first term on the right-hand side of \eqref{law-of-tv} 
approaches $\text{Var}(\theta_j|S_y)$.   Considering the second term, note
that \eqref{bounds} implies that 
$$
E(\theta_j|S_x)\in [E(\theta_j|S_y)- C^{-1}\epsilon,E(\theta_j|S_y)+ C^{-1}\epsilon].
$$
For fixed $n,\epsilon$, and $S_x\in\mathcal{B}_\epsilon$, we then have that $E(\theta_j|S_x)$ is a bounded random variable with a variance (conditional on $S_y$) that can be bounded by  $C^{-2}\epsilon^2$. Hence, its variance is decreasing to $0$ as $\epsilon\rightarrow 0$.  Hence 
\begin{align}
\lim_{\epsilon\rightarrow 0} \text{Var}_\epsilon(\theta_j) & =\text{Var}(\theta_j|S_y).  \label{moment2}
\end{align}

The interpretation of equations \eqref{moment1} and \eqref{moment2} is
that an ABC analysis using the idealized low-dimensional summary statistics $\widetilde{S}_j$ results
in ABC posterior mean and variance estimation 
similar to the exact partial posterior 
mean and variance for $\theta_j$ given $S_y$, at least as the tolerance goes to zero.

\begin{remark}  The above reasoning can be extended beyond a one-dimensional $\theta_j$. For example if $\theta_j=(\theta_{j1},\theta_{j2})^\top$, then the 
	idealized summary statistics 
	$$\widetilde{S}_j=(E(\theta_{j1}|S),E(\theta_{j2}|S),\text{Var}(\theta_{j1}|S),\text{Var}(\theta_{j2}|S),\text{Cov}(\theta_{j1},\theta_{j2}|S))^\top$$
	can be considered.
\end{remark}

\begin{remark} 
	We can also consider estimation of moments of higher than second order.  Consider the case of scalar $\theta_j$ and estimation of 
	moments up to third order.  Define
	$$\widetilde{S}_j=(E(\theta_j|S),\text{Var}(\theta_j|S),E((\theta_j-E(\theta_j|S))^3|S))^\top.$$
	For this choice of $\widetilde{S}_j$, the ABC posterior will accurately estimate
	moments up to third order for $\theta_j$ for the partial posterior given $S$.  
	The argument is similar to the case of second order moments, 
	but we can generalize the argument based on \eqref{law-of-tv} using the
	law of total cumulance \citep{brillinger69}.  The extension to moments
	of higher than third order is similar.
\end{remark}

Returning to the normal example in Section \ref{subsec:normal_example}, we consider running ABC for three different sets of summary statistics: (1) sample mean and standard deviation (sufficient), (2) marginal posterior mean (first moment) for either $\mu$ or $\phi$ and (3) marginal posterior mean and variance (first two moments) for either $\mu$ or $\phi$.  Since we do not use a conjugate prior here we do not have analytical expressions for the posterior moments.  As proxies we use the estimated posterior mode and variance obtained from a Laplace approximation of the posterior using a numerical optimiser where we optimise in terms of $\log \phi$ so that the optimiser can search over an unrestricted space.   

\begin{figure}[!htp]
	\centering
	\subfigure[Marginal posterior densities for $\mu$]{\includegraphics[scale=0.5]{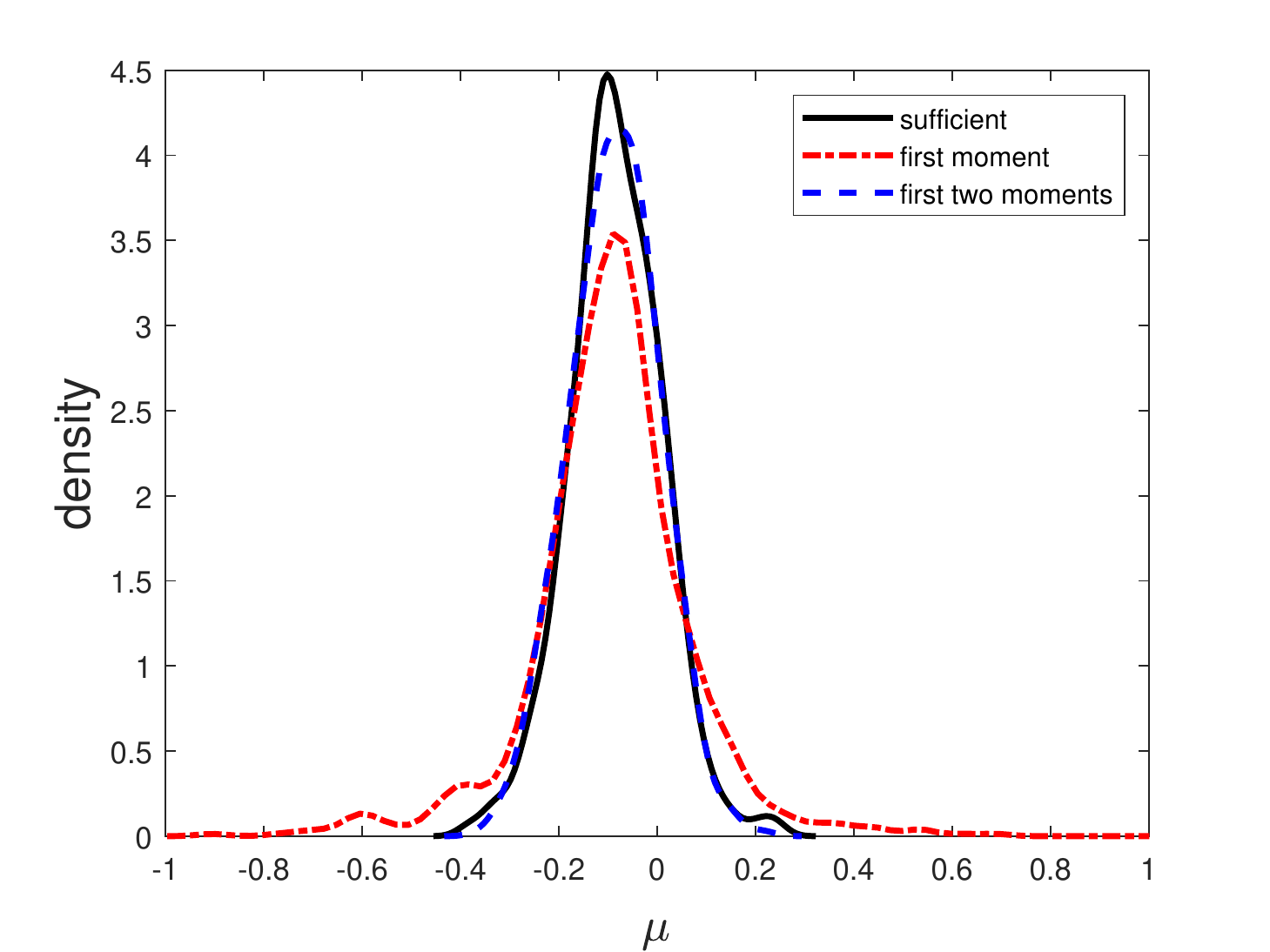}}
	\subfigure[Marginal posterior densities for $\phi$]{\includegraphics[scale=0.5]{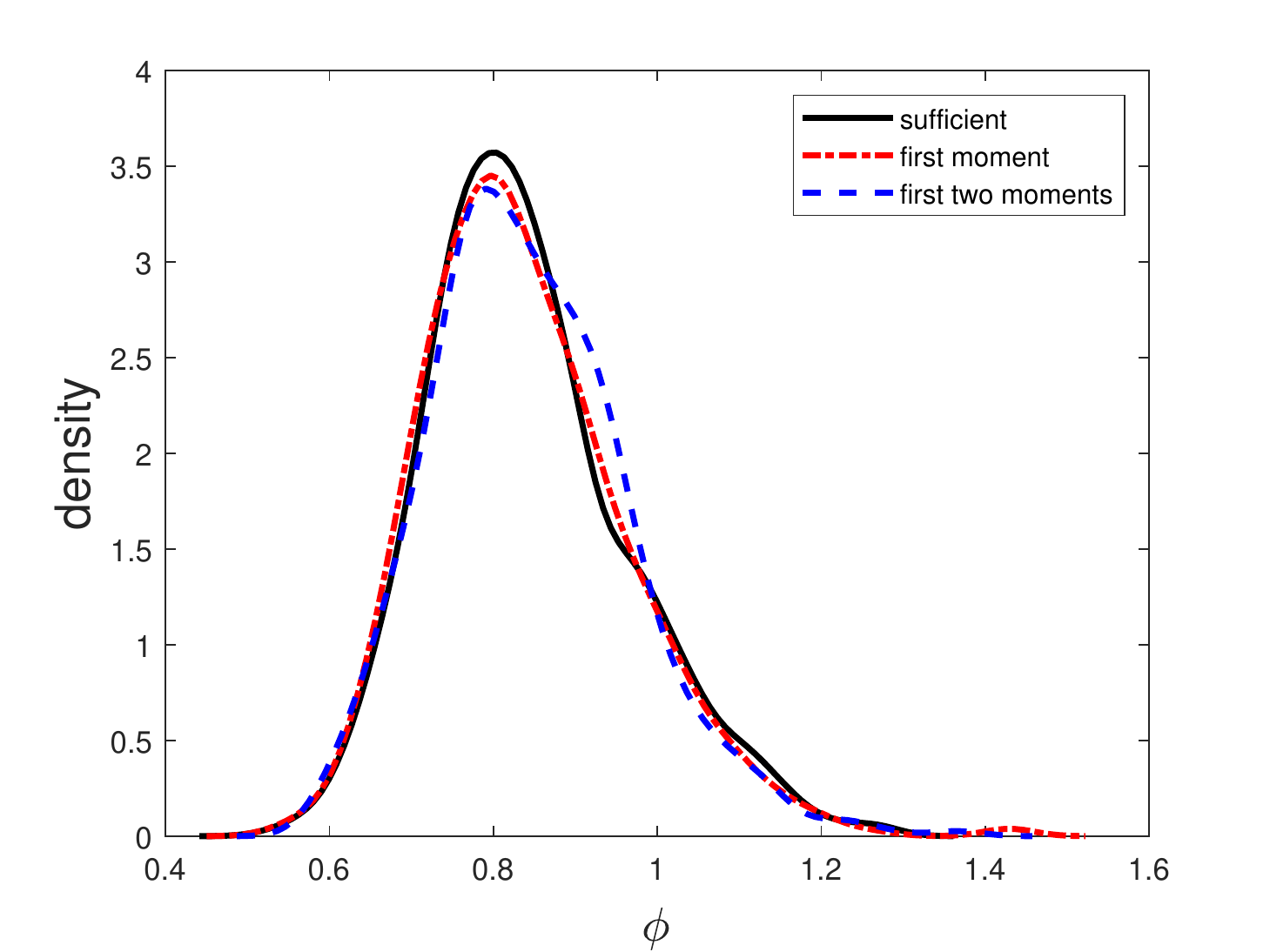}}
	\caption{Posteriors for toy normal example where we consider using marginal posterior means and variances as summary statistics.}
	\label{fig:results_toy_normal_meanvar}
\end{figure}

The results are shown in Figure \ref{fig:results_toy_normal_meanvar}.  It is evident that using only the posterior mean of $\mu$ as the summary statistic for ABC leads to a posterior approximation of $\mu$ where the location is well estimated but its variance is overestimated, leading to an inaccurate posterior approximation.  In contrast, using the first two posterior moments of $\mu$ as the summary statistic produces an accurate approximation as the ABC posterior of $\mu$ based on the sufficient statistics is well characterised by its first two moments.  For $\phi$, it seems that only the posterior mean of $\phi$ is required to produce an accurate approximation.

\subsection*{Practical Issues}

The idealized summary statistic vector $\widetilde{S}_j$ is not computable
in general, although we might wonder if we could approximate it using regression methods.  There is a precedent for such an approach in the ABC 
literature:  when approximating the full posterior, the   
widely used summary statistics of \citet{Fearnhead2012}  based on the 
posterior mean vector 
$E(\theta|S)$ for an initial set of candidate summary statistics
$S$, are usually implemented 
by using a regression approximation to the posterior mean estimated  
from pilot simulated data.  
However, in the case of marginal posterior approximations based on $\widetilde{S}_j$, we need to estimate the variance as a function of summary statistics
using regression, not just the mean, 
and this is much more difficult.  It should be kept in mind that the
approximation of the summary statistics, which could be computationally 
burdensome, would only be the first step in any ABC analysis.

Fortunately, we do not need to know the mean
and variance functions $E(\theta_j|S)$ and $\text{Var}(\theta_j|S)$.  
For simplicity, consider an indicator kernel in 
the ABC posterior \eqref{density-joint}, 
so that $K_\epsilon(\rho(S_y,S_x))=\mathbb{I}\{\rho(S_y,S_x) \le \epsilon\}$.  
Let's consider replacing the condition $\rho(S_y,S_x)\le\epsilon$ in
\eqref{density-joint} with
the condition $S_x\in A_y(\epsilon)$, where $A_y(\epsilon)$ satisfies
\begin{align}
& \sup_{S_x\in A_y(\epsilon)} |E(\theta_j|S_x)-E(\theta_j|S_y)|\le K\epsilon \;\;\text{ and }\;\;\sup_{S_x\in A_y(\epsilon)} |\text{Var}(\theta_j|S_x)-\text{Var}(\theta_j|S_y)|\le K\epsilon,\label{support-condition2}
\end{align}  
for some constant $K>0$.  
This leads to an approximation of the form \eqref{abc-posterior-subset}, 
where now
\begin{align*}
p_x(S_x|A_y(\epsilon)) & = \frac{\int \pi(\theta) p(S_x|\theta) \,d\theta \times \mathbb{I}(S_x\in A_y(\epsilon))}
{\int \int \pi(\theta) p(S_x|\theta) \,d\theta\, \mathbb{I}(S_x\in A_y(\epsilon))\,dS_x}.
\end{align*}
Examining our previous argument, we see that 
the condition \eqref{support-condition2} 
is all that we need to obtain accurate estimation of $E(\theta_j|S_y)$ and
$\text{Var}(\theta_j|S_y)$ for this generalized ABC approximation.  

An example of a situation where the condition \eqref{support-condition2} would be
satisfied is where we consider an ABC posterior conditioned on a summary
statistic $\breve{S}_j$ (possibly of higher dimension than $\widetilde{S}_j$) 
and such that $\widetilde{S}_j=f_j(\breve{S}_j)$, 
where $f_j(\cdot)$ is a continuous function.   
We do not need to know the function $f_j(\cdot)$;  
we just need to know that it exists.   
How should we choose $\breve{S}_j$?  We will not pursue here the issue
of how we can find a summary statistic $\breve{S}_j$ that is low-dimensional, 
but determines the marginal posterior mean and variance for $\theta_j$
given $S$, and strategies for doing this
will be pursued in future work.

Instead, in the main paper, we consider \eqref{support-condition2} with 
$$A_y(\epsilon)=\{S_x: \rho(S_y,S_x) \le \epsilon_0, \; \rho(S_{j,y},S_{j,x}) \le \epsilon_j\},$$
where $S_j$ is an initial choice of a low-dimensional summary statistic
informative about the marginal posterior location for $\theta_j$, and $\epsilon=(\epsilon_0,\epsilon_j)$, with $\epsilon_0$
and $\epsilon_j$ two tolerances with $\epsilon_0>\epsilon_j$.    
As an example of a choice of $S_j$, this could be constructed from 
the statistics of \citet{Fearnhead2012}, by extracting from the full set the estimated posterior 
mean for $\theta_j$.  For specific problems there may be other choices, which we highlight in the examples of Section \ref{sec:examples}.  We can see that
\begin{align}
\mathbb{I}\{S_x\in A_y(\epsilon)\} & =\mathbb{I}\{\rho(S_y,S_x) \le \epsilon_0\} \cdot \mathbb{I}\{\rho(S_{j,y},S_{j,x}) \le \epsilon_j\}.  \label{joint-matching}
\end{align}
Our use of this joint matching condition is informed by the intuition gained from our discussion of idealized summary statistics for marginal estimation, which
makes it clear that it is crucial to include information
about marginal posterior scale as well as location in the estimation
of posterior marginal distributions.

\end{document}